# THE ART OF FINDING THE OPTIMAL SCATTERING CENTER(S)[†]

Alexander V. Kildishev,[*,‡,¶] Karim Achouri,[§] and Daria Smirnova[∥]

‡*Elmore Family School of Electrical and Computer Engineering and Birck Nanotechnology Center, Purdue University, 1205 W State St, West Lafayette, IN 47907, USA*

¶*On sabbatical leave: ARC Centre of Excellence for Transformative Meta-Optical Systems (TMOS), Research School of Physics, Australian National University, Canberra, ACT 2601, Australia, and the Nanophotonics and Metrology Laboratory, Institute of Electrical and Microengineering, École Polytechnique Fédérale de Lausanne, Route Cantonale, 1015 Lausanne, Switzerland*

§*Nanophotonics and Metrology Laboratory, Institute of Electrical and Microengineering, École Polytechnique Fédérale de Lausanne, Route Cantonale, 1015 Lausanne, Switzerland.*

∥*ARC Centre of Excellence for Transformative Meta-Optical Systems (TMOS), Research School of Physics, Australian National University, Canberra, ACT 2601, Australia*

E-mail: kildishev@purdue.edu

Phone: +1 765 586 7435

**Abstract**

The efficient use of a multipole expansion of the far field for rapid numerical modeling and optimization of the optical response from ordered and disordered arrays of various structural elements is complicated by the ambiguity in choosing the ultimate expansion centers for individual scatterers. Since the multipolar decomposition depends on the position of the expansion center, the sets of multipoles are not unique. They may require constrained optimization to get the compact and most efficient spatial spectrum for each scatterer. We address this problem by finding the optimal scattering centers for which the spatial multipolar spectra become unique. We separately derive these optimal positions for the electric and magnetic parts by minimizing the norm of the poloidal electric and magnetic quadrupoles. Employing the long-wave approximation (LWA) ansatz, we verify the approach with theoretical discrete models and realistic scatterers. Expansion of the work beyond the LWA is possible, with a promise for faster and universal numerical schemes. We show that the optimal electric and magnetic scattering centers, in all cases, are not co-local with the centers of mass. The optimal multipoles, including the toroidal terms, are calculated for several structurally distinct scattering cases, and their utility for low-cost numerical schemes is discussed. The significant reduction of redundant computations through our optimized multipole descriptors will benefit emerging differentiable electromagnetic solvers, including the generalized T-matrix and MMP, which are critical for enabling machine learning-driven inverse design in photonics and optoelectronics. The approach could lead to faster convergence and better accuracy in numerical simulations of scattering behavior. The computational efficiency gains also directly translate into higher fidelity results. Analysis of the optically dispersive mutual positions of expansion centers based on our technique could improve the fundamental understanding of scattering and optical forces and torques. We also show that the number of optimal magnetic scattering centers can be linked to the multiplicity and topological metrics of the resonant modes excited within a given scatterer structure. This finding hints at potential connections to fundamental studies in optomechanics at the nanoscale, as well as quantum and topological photonics.

---

[†]FINDING THE OPTIMAL SCATTERING CENTER(S)

# Introduction

Cartesian and spherical multipole expansions[1–5] present helpful and compact mathematical descriptors of the physical fields and provide comprehensive insights into their complex interactions with the matter. Thus, the classical multipole theories of the Newtonian, electrostatic, and vector potentials[6] cover a diverse range of applications that include mathematical modeling in geomagnetism,[7] physical chemistry,[5] quantum biochemistry,[8] quantum electrodynamics,[4,5] non-linear optics[9] and metasurface modeling.[10–14] The compact multipole descriptors are also used in computational geometry and 3D image recognition[15,16] and include applications to static and dynamic models of brain[17] and heart.[18,19] Such a diverse variety of viable applications also inspired 3D artistic renditions, such as the "Geo Face Distributor" sculpture by James Angus (The National Portrait Galery, Canberra, Australia).

The truncated spatial multipolar spectra of the emitted and scattered fields enable efficient and scalable numerical schemes that approximate solutions to acoustic, quantum-mechanical and electromagnetic problems with controlled accuracy.[8,20]

The far-field interactions of the dominant multipoles in the linear and nonlinear regimes lead to the novel supercavity paradigms.[21,22] Associated with the bound states in the continuum,[23] these recent demonstrations of high-Q resonant systems can be easily explained, designed, and optimized employing the multipole content of the scattered fields.[24,25] The optimized multipolar models of optical radiating and scattering sources are also critical for a comprehensive understanding of their topological properties, connecting the multipolar spectra with the line singularities of Hopf indices and the Bloch modes.[26–28]

The broadly adopted modern multiscale and multiphysics approaches require reducing the computational cost of their predictive models that enable combining other physics with complex light-matter interactions. Differentiable electromagnetic solvers, including the generalized multipole approach,[29,30] constitute an advantageous class of numerical engines for AI-driven frameworks for the inverse design of photonic structures. The efficient use of compact multipole descriptors in such frameworks is complicated by the ambiguity in choosing ultimate expansion centers for individual sources or scatterers. Since the multipolar decomposition depends on the position of the expansion center, the sets of multipoles are not unique and may require constrained optimization to get the ultimate spatial spectrum.

Such a non-uniqueness has been well understood in the problem of the scalar potential fields. Historical examples include the position of the ultimate geomagnetic center due to Kelvin[31] and Schmidt,[32,33] which remain instrumental in modern geophysics[34] and spacecraft engineering.[35,36]

Exploring the scalar electric potential of the heart, represented by a dipole plus quadrupole, Geselowitz[18] defined an optimum dipole location as the point where the dipole term alone gives the best least square fit to the potential. Bentley and Stone independently provided systematic studies of the optimal positioning of expansion centers for molecular electrostatic potentials, offering key insights into obtaining rapidly convergent multipole expansions (see,[37][38] and references within). These works laid the foundations for later applications of multipole methods in computational chemistry and physics. Finding the ultimate electrostatic center in computational chemistry is typically concerned with a compact and unique distribution of charges and assumes a monopole term,[5,39,40] with the ultimate position of the molecular electrostatic center comprehensively presented by Platt and Silverman.[41]

Retrieving the optimal multipole centers beyond the quasistatic approximation has not been explored in classical or quantum electrodynamics and can enable new advanced theoretical concepts and efficient numerical schemes. We expand this problem beyond the electro- and magnetostatics and find the optimal multipole centers for which the dynamic spatial multipolar spectra become unique. Employing the truncated long-wave approximation (LWA) ansatz,[10,42–45] we demonstrate the approach by deriving the ultimate positions for the distinct origins of electric and magnetic multipole expansions by minimizing the norm of the poloidal electric and magnetic terms. We also confirm an intuitively apparent notion that, in many practical cases, the optimal electric and magnetic centers do not coincide with the center of mass. The optimal multipoles are calculated for several structurally distinct scattering cases, and their utility for low-cost numerical schemes is discussed.

## Problem statement and outline

As we describe a scatterer or source with a truncated set of multipoles (e.g., electric and magnetic dipoles $\{\mathbf{p}, \mathbf{m}\}$, electric and magnetic quadrupoles $\{\mathbf{Q}_\text{e}, \mathbf{Q}_m\}$, etc.), we always want to find the best approximation to the fields, measured or computed over a basis boundary $S$.

If we change the relative position of the scatterer and $S$, then the multipole expansion is translated to a new origin $O \to O'$. Since for any general scatterer (or source), its multipole expansion over the basis boundary $S$ depends on the position of $S$ relative to the origin $O'$, a new set of multipoles appears every time we move $O \to O'$. Therefore, it is desirable to find *the optimal scattering center position* $O'$ relative to the expansion boundary $S$ and a matching, optimal set of multipoles, provided that the initial (nonoptimal) multipole expansion is known either from the numerical model or experiment.

As we show in this study, the optimal center typically does not match the center of mass (CM) computed for the scattering or emitting domain. The mismatch between the material's optical polarizability and its mass density, along with the existence of negative susceptibility, suggest that the notion of the ultimate center differs fundamentally from that of the center of mass. Moreover, the ultimate scattering center depends on the photon energy and its position exhibits optical dispersion. We argue, therefore, that although a common practice of choosing the multipole expansion center at the center of mass (see, e.g., Ref.[46]) does not have a solid ground, it could serve as a good starting point for finding the optimal scattering centers.

# Results and discussion

## Nomenclature

We start with defining the position vector, and the Cartesian basis,

$$\mathbf{r} = \begin{pmatrix} r_x \\ r_y \\ r_z \end{pmatrix}; \ \hat{\mathbf{x}} = \begin{pmatrix} 1 \\ 0 \\ 0 \end{pmatrix}; \ \hat{\mathbf{y}} = \begin{pmatrix} 0 \\ 1 \\ 0 \end{pmatrix}; \ \hat{\mathbf{z}} = \begin{pmatrix} 0 \\ 0 \\ 1 \end{pmatrix}. \quad (1)$$

We also include a general vector product notation,

$$\mathbf{A} \times \mathbf{v} = \begin{pmatrix} \mathbf{a}_1 \times \mathbf{v} \\ \mathbf{a}_2 \times \mathbf{v} \\ \mathbf{a}_3 \times \mathbf{v} \end{pmatrix},$$

for any $3 \times 3$ tensor $\mathbf{A} = (\mathbf{a}_1, \mathbf{a}_2, \mathbf{a}_3)^{\mathrm{T}}$, and a vector $\mathbf{v} = (v_1, v_2, v_3)^{\mathrm{T}}$, along with a dyadic product of two vectors, $\mathbf{v} \otimes \mathbf{v}$. We also define a general symmetrizing operator, $\mathbf{A}^{\mathrm{S}} = \mathbf{A} + \mathbf{A}^{\mathrm{T}}$ and the auxiliary dyads $\hat{\mathbf{X}}, \hat{\mathbf{Y}}, \hat{\mathbf{Z}}$ and $\hat{\mathbf{I}}$,

$$\hat{\mathbf{I}} = (\hat{\mathbf{x}}, \hat{\mathbf{y}}, \hat{\mathbf{z}}); \ \hat{\mathbf{X}} = \hat{\mathbf{x}} \otimes \hat{\mathbf{x}}, \ \hat{\mathbf{Y}} = \hat{\mathbf{y}} \otimes \hat{\mathbf{y}}, \ \hat{\mathbf{Z}} = \hat{\mathbf{z}} \otimes \hat{\mathbf{z}}. \quad (2)$$

The standard constant $s = -i\omega$ is assumed in the study.

### The LWA dyadic form of the E-field

Employing a known definition,[45] we can write the E-field generated by multipoles at any far-field point $\mathbf{r}$ in a dyadic form:

$$\mathbf{E} = \kappa \left[ \left( \hat{\mathbf{r}} \times \mathbf{p} + \frac{1}{c}\mathbf{m} \right) - ik \left( \hat{\mathbf{r}} \times (\mathbf{Q}_{\mathrm{e}} \cdot \hat{\mathbf{r}}) + \frac{1}{c}\mathbf{Q}_{\mathrm{m}} \cdot \hat{\mathbf{r}} \right) \right] \times \hat{\mathbf{r}}, \quad (3)$$

with $\kappa = \frac{k^2 \mathrm{e}^{ikr}}{4\pi\varepsilon_0 r}$.

Finding an optimal scattering center $O'$, assuming a translation $\mathbf{r}' = \mathbf{r} + \mathbf{d}$ relative to the initial origin $O$ with displacement **d** in an arbitrary direction (see Fig. 1), requires the knowledge of few first multipole terms and their dependence on the offset **d**. These dependencies, derived in the Supporting Information, are summarized below.

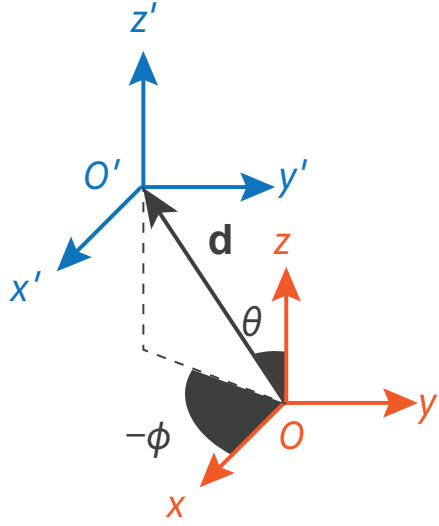

Figure 1: **Translation of reference origin $O \to O'$ by a vector d.**

## Summary of LWA Multipoles

The explicit formulae for the first two orders of unshifted poloidal Cartesian multipoles (at the origin $O$) are given in Table 1, where all integrals are normalized to match (3).

Table 1: Unshifted poloidal LWA multipoles (first two orders)

| $n$ | Electric multipoles | Magnetic multipoles |
|---|---|---|
| 1 | $\mathbf{p}_0 = \frac{1}{s} \int dV \, \mathbf{J}$ | $\mathbf{m}_0 = \frac{1}{2} \int dV \, (\mathbf{r} \times \mathbf{J})$ |
| 2 | $\mathbf{Q}_{e1} = \frac{1}{2s} \int dV (\mathbf{r} \otimes \mathbf{J})$ | $\mathbf{Q}_{m1} = \frac{1}{6} \int dV \, [\mathbf{r} \otimes (\mathbf{r} \times \mathbf{J})]$ |
|   | $\mathbf{Q}_{pe} = (\mathbf{Q}_{e1})^S - \frac{2}{3} \hat{\mathbf{I}} \, \mathrm{Tr}(\mathbf{Q}_{e1})$ | $\mathbf{Q}_{pm} = (\mathbf{Q}_{m1})^S$ |

Note that the electric quadrupole $\mathbf{Q}_e$ is both symmetrized and detraced, whereas the magnetic quadrupole $\mathbf{Q}_m$ does not require detracing since $\mathrm{Tr}(\mathbf{Q}_{m1}) = \mathbf{r} \cdot (\mathbf{r} \times \mathbf{J}) \equiv 0$.

Next, Table 2 displays the first two orders of the poloidal moments, shifted to the origin

$O'$). Then, Table 3 gives the first two orders of the unshifted toroidal moments, normalized

Table 2: Shifted poloidal LWA multipoles (first two orders)

| $n$ | Electric multipoles | Magnetic multipoles |
|---|---|---|
| 1 | $\mathbf{p}'_0 \equiv \mathbf{p}_0$ | $\mathbf{m}'_0 = \mathbf{m}_0 + \frac{s}{2}\mathbf{d}\times\mathbf{p}_0$ |
|   | $\mathbf{Q}'_{e1} = \mathbf{Q}_{e1} + \frac{1}{2}\mathbf{d}\otimes\mathbf{p}_0$ | $\mathbf{Q}'_{m1} = \mathbf{Q}_{m1}+$ |
|   |  | $+\frac{1}{3}\left[\mathbf{d}\otimes(\mathbf{m}_0 - \frac{s}{2}\mathbf{d}\times\mathbf{p}_0) - s\,\mathbf{Q}_{e1}\times\mathbf{d}\right]$ |
| 2 | $\mathbf{Q}'_{pe} = (\mathbf{Q}'_{e1})^S - \frac{2}{3}\hat{\mathbf{I}}\operatorname{Tr}(\mathbf{Q}'_{e1})$ | $\mathbf{Q}'_{pm} = (\mathbf{Q}'_{m1})^S$ |
|   | $\mathbf{Q}'_{pe} = \mathbf{Q}_{pe}+$ | $\mathbf{Q}'_{pm} = \mathbf{Q}_{pm}+$ |
|   | $+\frac{1}{2}(\mathbf{d}\otimes\mathbf{p}_0)^S - \frac{1}{3}\hat{\mathbf{I}}(\mathbf{d}\cdot\mathbf{p}_0)$ | $+\frac{1}{3}\left[\mathbf{d}\otimes(\mathbf{m}_0 - \frac{s}{2}\mathbf{d}\times\mathbf{p}_0) - s\,\mathbf{Q}_{e1}\times\mathbf{d}\right]^S$ |

to match the electric field in (3), and where the coefficient $a$ denotes $a = -\frac{k^2}{10s}$. Finally, the first two orders of the shifted toroidal moments are listed in Table 4.

Table 3: Unshifted toroidal LWA multipoles (first two orders)

| $n$ | Electric multipoles | Magnetic multipoles |
|---|---|---|
|   | $\mathbf{T}_e = a\int dV\,[\mathbf{J}\otimes(\mathbf{r}\otimes\mathbf{r})]$ |  |
|   | $\mathbf{p}_{t1} = \mathbf{p}_{t2} + \sum_{\hat{\mathbf{q}}}^{\{\hat{\mathbf{x}},\hat{\mathbf{y}},\hat{\mathbf{z}}\}}\hat{\mathbf{q}}\cdot(\hat{\mathbf{q}}\cdot\mathbf{T}_e)$ | $\mathbf{T}_m = \frac{as}{2}\int dV\,(\mathbf{r}\otimes\mathbf{r})\otimes(\mathbf{J}\times\mathbf{r})$ |
| 1 | $\mathbf{p}_{t2} = -\sum_{\hat{\mathbf{q}}}^{\{\hat{\mathbf{x}},\hat{\mathbf{y}},\hat{\mathbf{z}}\}}(\mathbf{T}_e\cdot\hat{\mathbf{q}})\cdot\hat{\mathbf{q}}$ |  |
|   | $\mathbf{p}_t = \mathbf{p}_{t1} + \mathbf{p}_{t2} =$ | $\mathbf{m}_t = \sum_{\hat{\mathbf{q}}}^{\{\hat{\mathbf{x}},\hat{\mathbf{y}},\hat{\mathbf{z}}\}}\operatorname{Tr}(\mathbf{T}_m\cdot\hat{\mathbf{q}})\,\hat{\mathbf{q}}$ |
|   | $= \sum_{\hat{\mathbf{q}}}^{\{\hat{\mathbf{x}},\hat{\mathbf{y}},\hat{\mathbf{z}}\}}[\hat{\mathbf{q}}\cdot(\hat{\mathbf{q}}\cdot\mathbf{T}_e) - 2(\mathbf{T}_e\cdot\hat{\mathbf{q}})\cdot\hat{\mathbf{q}}]$ |  |
|   | $\mathbf{Q}_{e2} = \frac{10}{21}a\int dV\,(\mathbf{r}\otimes\mathbf{r})(\mathbf{r}\cdot\mathbf{J})$ | $\mathbf{Q}_{m2} = -\frac{as}{84}\int dV\,r^2\mathbf{r}\otimes\mathbf{r}\times\mathbf{J}$ |
| 2 | $\mathbf{Q}_{e3} = -\frac{25}{42}a\int dV\,r^2(\mathbf{r}\otimes\mathbf{J})$ | $\mathbf{Q}_{tm} = (\mathbf{Q}_{m2})^S$ |
|   | $\mathbf{Q}_{te} = \mathbf{Q}_{e2} + \frac{1}{2}\hat{\mathbf{I}}\operatorname{Tr}(\mathbf{Q}_{e2}) + (\mathbf{Q}_{e3})^S$ |  |

# Optimal Scattering Centers

## Optimal Electric Scattering Center (OSC$^e$)

Employing the components of $\mathbf{Q}_{pe}$ and $\mathbf{p}_0$ for minimizing the norm of the shifted poloidal electric quadrupole, $\min\left[\operatorname{Tr}(\mathbf{Q}'_{pe}\cdot\mathbf{Q}'^{\dagger}_{pe})\right] \equiv \min\left[\sum|Q'_{pe\{i,j\}}|^2\right]$, we can get the optimal offset

Table 4: Shifted toroidal LWA multipoles (first two orders)

| $n$ | Electric multipoles | Magnetic multipoles |
|---|---|---|
| 1 | $\mathbf{p}'_{t1} = \mathbf{p}_{t1} + 2a\,\mathbf{d} \times \mathbf{m}'_0 +$ <br> $+ 2as\left[\text{Tr}(\mathbf{Q}_{e1})\mathbf{d} - \mathbf{d}\cdot\mathbf{Q}_{e1}\right]$ <br> $\mathbf{p}'_{t2} = \mathbf{p}_{t2} - as\left(d^2\,\mathbf{p}_0 + 4\mathbf{d}\cdot\mathbf{Q}_{e1}\right)$ <br> $\mathbf{p}'_t = \mathbf{p}'_{t1} + \mathbf{p}'_{t2}$ | $\mathbf{m}'_t = \mathbf{m}_t -$ <br> $-as(d^2\mathbf{m}'_0 + 6\mathbf{d}\cdot\mathbf{Q}_{m1}) +$ <br> $+\frac{s}{2}\mathbf{d}\times(\mathbf{p}_{t2} - 4as\mathbf{d}\cdot\mathbf{Q}_{e1})$ |
| 2 | $\mathbf{Q}'_{e2} = \mathbf{Q}_{e2} +$ <br> $+\frac{10}{21}\left[\mathbf{d}\cdot\mathbf{T}_e + (\mathbf{d}\otimes(\mathbf{p}_{t1} - \mathbf{p}_{t2}))^{\text{S}}\right]$ <br> $+\frac{20as}{21}\left[(\mathbf{Q}_{e1}\cdot(\mathbf{d}\otimes\mathbf{d}))^{\text{S}}\right] +$ <br> $+\frac{20as}{21}\left[\text{Tr}(\mathbf{Q}_{e1})\,\mathbf{d}\otimes\mathbf{d}\right]$ <br> $\mathbf{Q}'_{e3} = \mathbf{Q}_{e3} -$ <br> $-\frac{25}{21}\mathbf{T}_e\cdot\mathbf{d} + \frac{25}{42}\mathbf{d}\otimes\mathbf{p}_{t2} -$ <br> $-\frac{25as}{21}\left(2(\mathbf{d}\otimes\mathbf{d})\cdot\mathbf{Q}_{e1} + d^2\mathbf{Q}'_{e1}\right)$ <br> $\mathbf{Q}'_{te} = \mathbf{Q}'_{e2} + \frac{1}{2}\hat{\mathbf{I}}\text{Tr}(\mathbf{Q}'_{e2}) + (\mathbf{Q}'_{e3})^{\text{S}}$ | $\mathbf{Q}'_{m2} = \mathbf{Q}_{m2} +$ <br> $+\frac{1}{21}\mathbf{d}\cdot(\mathbf{T}_m + \frac{s}{2}[\mathbf{T}_e]^{\text{M}} \times \mathbf{d})$ <br> $-\frac{as}{7}\mathbf{d}\otimes\left(\mathbf{d}\cdot\mathbf{Q}_{m1} + \frac{s}{3}\mathbf{d}\times(\mathbf{d}\cdot\mathbf{Q}_{e1})\right)$ <br> $+\frac{1}{42}\mathbf{d}\otimes\left(\frac{s}{2}\mathbf{d}\times\mathbf{p}_{t2} + \mathbf{m}_t\right)$ <br> $-\frac{asd^2}{14}\mathbf{Q}'_{pm} - \frac{s}{50}\mathbf{Q}_{e3}\times\mathbf{d}$ <br> $\mathbf{Q}'_{tm} = (\mathbf{Q}'_{m2})^{\text{S}}$ |

$\mathbf{d}_e$ for the poloidal electric multipoles.

Upon translation $\mathbf{d}$, the multipole terms of $\mathbf{Q}'_{pe}$ are related to $\mathbf{Q}_{pe}$ with $\mathbf{p}_0$ through, $\mathbf{Q}'_{pe} = \mathbf{Q}_{pe} + \boldsymbol{\Delta}_{pe}$, where $\boldsymbol{\Delta}_{pe} = \frac{1}{2}(\mathbf{d}\otimes\mathbf{p}_0)^{\text{S}} - \frac{1}{3}\hat{\mathbf{I}}(\mathbf{d}\cdot\mathbf{p}_0)$.

We rewrite $\text{Tr}\left[\mathbf{Q}'_{pe}\cdot\mathbf{Q}'^{\dagger}_{pe}\right]$ as a sum of three traces $(t_l)$ of order $l = 0, 1, 2$,

$$\text{Tr}\left[\mathbf{Q}'_{pe}\cdot\mathbf{Q}'^{\dagger}_{pe}\right] \equiv \text{Tr}\left[(\mathbf{Q}_{pe} + \boldsymbol{\Delta}_{pe})\cdot(\mathbf{Q}_{pe} + \boldsymbol{\Delta}_{pe})^{\dagger}\right] \equiv \sum_{l}^{0,1,2} t_l, \quad (4)$$

where

$$t_0 = \text{Tr}\left[\mathbf{Q}_{pe}\cdot\mathbf{Q}^*_{pe}\right]; \quad t_1 = \text{Tr}\left[\mathbf{Q}_{pe}\cdot\boldsymbol{\Delta}^*_{pe} + \boldsymbol{\Delta}_{pe}\cdot\mathbf{Q}^*_{pe}\right]; \quad t_2 = \text{Tr}\left[\boldsymbol{\Delta}_{pe}\cdot\boldsymbol{\Delta}^*_{pe}\right]. \quad (5)$$

First, we define the first-order trace $t_1$ as $t_1 = \mathbf{b}\cdot\mathbf{d}$ with $\mathbf{b} = \sum_{q}^{x,y,z} b_q\hat{\mathbf{q}}$, so that $\nabla_{\mathbf{d}}(t_1) =$ **b**, where the vector **b** is given by

$$\mathbf{b} = 2\left[\Re(\mathbf{Q}_{\text{pe}}^* \cdot \mathbf{p}_0) - \frac{1}{3}\Re[\mathbf{p}_0 \text{Tr}(\mathbf{Q}_{\text{pe}}^*)]\right]. \tag{6}$$

Then, in a similar fashion, we use tensor $\mathbf{A} = (\mathbf{a}_x, \mathbf{a}_x, \mathbf{a}_z)$ to rewrite the gradient of $t_2$ as, $\nabla_{\mathbf{d}}(t_2) = \mathbf{A} \cdot \mathbf{d}$, where the tensor $\mathbf{A}$ reads

$$\mathbf{A} = \frac{1}{3}\Re(\mathbf{p}_0^* \otimes \mathbf{p}_0) + p_0^2 \hat{\mathbf{I}}. \tag{7}$$

Finally, we ignore the vanishing gradient of the trace $t_0$ and directly solve a linear equation $\nabla_{\mathbf{d}}(t_2 + t_1) = \mathbf{0}$, i.e., $\mathbf{A} \cdot \mathbf{d}_e + \mathbf{b} = \mathbf{0}$. The latter could have been solved by inverting $\mathbf{A}$ so that, $\mathbf{d}_e = -\mathbf{A}^{-1} \cdot \mathbf{b}$. Our alternative approach utilizes a normalized and shifted tensor $\mathbf{S} = p_0^{-2}\mathbf{A} - \hat{\mathbf{I}} = \frac{1}{6p_0^2}(\mathbf{p}_0^* \otimes \mathbf{p}_0)^S$ and employs its determinant $|\mathbf{S}|$, trace $\text{Tr}(\mathbf{S})$, adjugate $\text{Adj}(\mathbf{S})$, and the trace of adjugate $\text{Tr}[\text{Adj}(\mathbf{S})]$.

We use an *ad hoc* lemma (see Supporting Information) to write the optimal electric scattering center as $\mathbf{d}_e = \frac{\mathbf{S} - [\hat{\mathbf{I}} + \hat{\mathbf{I}}\text{Tr}(\mathbf{S}) + \text{Adj}(\mathbf{S})]}{1 + \text{Tr}(\mathbf{S}) + |\mathbf{S}| + \text{Tr}[\text{Adj}(\mathbf{S})]} \cdot (p_0^{-2}\mathbf{b})$. We implement the above splitting to take advantage of apparent simplifications ($\text{Tr}(\mathbf{S}) = 1/3$ and $|\mathbf{S}| = 0$), while we keep $\text{Adj}(\mathbf{S})$ that, for a singular $\mathbf{S}$, satisfies $\mathbf{S} \cdot \text{Adj}(\mathbf{S}) \equiv |\mathbf{S}|\hat{\mathbf{I}} = \mathbf{0}$). Then, $\mathbf{d}_e$ yields

$$\mathbf{d}_e = -\frac{3\left[\text{Adj}(\mathbf{S}) - \mathbf{S}\right] + 4\hat{\mathbf{I}}}{3\text{Tr}[\text{Adj}(\mathbf{S})] + 4} \cdot (p_0^{-2}\mathbf{b}). \tag{8}$$

**OSC$^e$: the Axisymmetric Case**

Equation (8) may be simplified by considering the special case of a source or scatterer with zero-order rotational symmetry. In this case, we only have the offset $d_{\text{ez}}$ confined to the axis of symmetry, with a simpler linear equation $A_{\text{zz}}d_{\text{ez}} + b_{\text{z}} = 0$, where $A_{\text{zz}} = |\mathbf{p}_0|^2 + |p_{0z}|^2$ and $b_{\text{z}} = 2\Re\left[p_{0x}^*(\mathbf{Q}_{\text{pe}})_{\text{xz}} + p_{0y}^*(\mathbf{Q}_{\text{pe}})_{\text{yz}} + p_{0z}^*(\mathbf{Q}_{\text{pe}})_{\text{zz}}\right]$, which takes into account that $\mathbf{Q}_{\text{pe}}$ is symmetric and traceless. So then, the optimal axial offset is

$$d_{\text{ez}} = -\frac{b_{\text{z}}}{A_{\text{zz}}} = -\frac{\Re\left[2\left(p_{0\text{x}}^*(\mathbf{Q}_{\text{pe}})_{\text{xz}} + p_{0y}^*(\mathbf{Q}_{\text{pe}})_{\text{yz}}\right) + 3p_{0z}^*(\mathbf{Q}_{\text{pe}})_{\text{zz}}\right]}{|\mathbf{p}_0|^2 + |p_{0\text{z}}|^2}, \quad (9)$$

with $\mathbf{p}_0 = \hat{\mathbf{q}}p_{0x} + \hat{\mathbf{y}}p_{0y} + \hat{\mathbf{z}}p_{0z}$; apparently, the offset is defined by six poloidal multipolar terms taken from $\mathbf{Q}_{\text{pe}}$ and $\mathbf{p}_0$.

Equations (8) and (9) define the electric OSC in the general and axisymmetric case, respectively, and are the main results of this section. Obtaining the optimal reference position for electric multipoles ($d_{\text{e}}$), using (8) requires solving a linear equation that we reduce analytically. In contrast to the electric center, getting the magnetic OSC requires solving a non-linear equation, as discussed in the next section.

## Optimal Magnetic Scattering Center (OSC$^{\text{m}}$)

The optimal electric scattering center (OSC$^{\text{e}}$, $O'_{\text{e}}$) offers an ultimate position of the electric part of a multipole expansion, minimizing the poloidal quadrupole content with a relatively simple analytical expression. The next question is whether we would need another translation $\mathbf{d}_{\text{m}} : O'_{\text{e}} \to O'_{\text{m}}$ for a new expansion center for magnetic terms that would minimize the poloidal magnetic quadrupole is yet to be answered. Several points must be considered before such a position $O'_{\text{m}}$ is derived.

We recapitulate here that, for any symmetric and traceless magnetic quadrupole tensor, an integral over the encapsulating sphere $S$ gives

$$\begin{aligned}
\frac{5}{4\pi}\int \mathrm{d}S\,|(\mathbf{Q}_{\text{pm}}\cdot\hat{\mathbf{r}})\times\hat{\mathbf{r}}|\cdot\left[(\mathbf{Q}_{\text{pm}}^\dagger\cdot\hat{\mathbf{r}})\times\hat{\mathbf{r}}\right] \\
\equiv \operatorname{Tr}\left(\mathbf{Q}_{\text{pm}}^\dagger\cdot\mathbf{Q}_{\text{pm}}\right) \equiv \sum_{i,j}|(\mathbf{Q}_{\text{pm}})_{i,j}|^2 \\
\equiv |Q_{xx}|^2 + |Q_{yy}|^2 + |Q_{zz}|^2 + 2(|Q_{xy}|^2 + |Q_{yz}|^2 + |Q_{zx}|^2).
\end{aligned} \quad (10)$$

In contrast with the OSC$^{\text{e}}$, the components of $\mathbf{Q}_{\text{pe}}$, $\mathbf{m}_0$, and $\mathbf{p}_0$ are required for minimizing the norm of the shifted poloidal magnetic quadrupole, $\min \left[\text{Tr}(\mathbf{Q}'_{\text{pm}} \cdot \mathbf{Q}'^{\dagger}_{\text{pm}})\right] \equiv \min \left[\sum |Q'_{\text{pm}\{i,j\}}|^2\right]$ and retrieving the optimal offset $\mathbf{d}_{\text{m}}$ for the OSC$^{\text{m}}$.

We start by noting that upon translation $\mathbf{d}$, $\mathbf{Q}'_{\text{pm}}$ is related to $\mathbf{Q}_{\text{pm}}$ with $\mathbf{p}_0, \mathbf{Q}_{\text{e1}}$, and $\mathbf{m}$ through $\mathbf{Q}'_{\text{pm}} = \mathbf{Q}_{\text{pm}} + \mathbf{\Delta}_{\text{pm}}$, where $\mathbf{\Delta}_{\text{pm}} = \frac{1}{3}\left[\mathbf{d} \otimes (\mathbf{m} - \frac{s}{2}\mathbf{d} \times \mathbf{p}_0) - s\,\mathbf{Q}_{\text{e1}} \times \mathbf{d}\right]^{\text{S}}$. Then, we rewrite $\text{Tr}\left[\mathbf{Q}'_{\text{pm}} \cdot \mathbf{Q}'^{\dagger}_{\text{pm}}\right]$ as a combination of three traces ($t_l$) of order $l = 0, 2, 4$ as

$$\text{Tr}\left[\mathbf{Q}'_{\text{pm}} \cdot \mathbf{Q}'^{\dagger}_{\text{pm}}\right] \equiv \text{Tr}\left[(\mathbf{Q}_{\text{pm}} + \mathbf{\Delta}_{\text{pm}}) \cdot (\mathbf{Q}_{\text{pm}} + \mathbf{\Delta}_{\text{pm}})^{\dagger}\right] \equiv \sum_l^{0,2,4} t_l, \tag{11}$$

where

$$t_0 = \text{Tr}\left[\mathbf{Q}_{\text{pm}} \cdot \mathbf{Q}^*_{\text{pm}}\right]; \; t_2 = \text{Tr}\left[\mathbf{Q}_{\text{pm}} \cdot \mathbf{\Delta}^*_{\text{pm}} + \mathbf{\Delta}_{\text{pm}} \cdot \mathbf{Q}^*_{\text{pm}}\right]; \; t_4 = \text{Tr}\left[\mathbf{\Delta}_{\text{pm}} \cdot \mathbf{\Delta}^*_{\text{pm}}\right]. \tag{12}$$

**Spatial Derivatives of the Traces**

We ignore the zero-order trace $t_0$ and consider a new definition for the symmetrized term ($\mathbf{\Delta}_{\text{pm}}$) that modifies the shifted poloidal magnetic quadrupole, $\mathbf{\Delta}_{\text{pm}} = \frac{1}{3}\left[\mathbf{N} \cdot \mathbf{d} + \frac{s}{2}(\mathbf{A}_{\text{s}} \cdot \mathbf{d}) \cdot \mathbf{d}\right]$, where $\mathbf{N} = (\mathbf{M} - s\mathbf{Q}_{\text{se}})$. The auxiliary symmetric tensors ($\mathbf{M}$ and $\mathbf{Q}_{\text{se}}$) of rank three and dimensions $3 \times 3 \times 3$, are defined as $\mathbf{M} = (\mathbf{m}_0 \otimes \hat{\mathbf{x}}, \mathbf{m}_0 \otimes \hat{\mathbf{y}}, \mathbf{m}_0 \otimes \hat{\mathbf{z}})^{\text{S}}$ and $\mathbf{Q}_{\text{se}} = (\mathbf{Q}_{\text{e1}} \times \hat{\mathbf{x}}, \mathbf{Q}_{\text{e1}} \times \hat{\mathbf{y}}, \mathbf{Q}_{\text{e1}} \times \hat{\mathbf{z}})^{\text{S}}$, while the higher-dimension $(3 \times 3 \times 3 \times 3)$ symmetric tensor $\mathbf{A}_{\text{s}}$ of rank four, is given by a nested array as

$$\mathbf{A}_{\text{s}} = \begin{pmatrix} (\mathbf{A}_{\text{xx}}) & (\mathbf{A}_{\text{xy}}) & (\mathbf{A}_{\text{zx}}) \\ (\mathbf{A}_{\text{xy}}) & (\mathbf{A}_{\text{yy}}) & (\mathbf{A}_{\text{yz}}) \\ (\mathbf{A}_{\text{zx}}) & (\mathbf{A}_{\text{yz}}) & (\mathbf{A}_{\text{zz}}) \end{pmatrix},$$

with $\mathbf{A}_{\text{ij}} = \hat{\mathbf{j}} \otimes \mathbf{p}_0 \times \hat{\mathbf{i}} + \hat{\mathbf{i}} \otimes \mathbf{p}_0 \times \hat{\mathbf{j}}$, $\forall \{i,j\} \in \{x,y,z\}$ and matching $\{\hat{\mathbf{i}}, \hat{\mathbf{j}}\} \in \{\hat{\mathbf{x}}, \hat{\mathbf{y}}, \hat{\mathbf{z}}\}$. For brevity, the explicit nested-matrix form of the tensor is shown in the Supporting Information.

To get the optimal solution we use (12) and derive $\partial_q t_2$ and $\partial_q t_4$ $\forall$ $q = \{x, y, z\}$, which gives

$$\partial_q t_2 = 2\Re\left(\text{Tr}\left[\mathbf{Q}_{\text{pm}}^* \cdot \partial_q \mathbf{\Delta}_{\text{pm}}\right]\right), \tag{13}$$

$$\partial_q t_4 = 2\Re\left(\text{Tr}\left[\mathbf{\Delta}_{\text{pm}}^* \cdot \partial_q \mathbf{\Delta}_{\text{pm}}\right]\right), \tag{14}$$

with

$$\partial_q \mathbf{\Delta}_{\text{pm}} = \frac{1}{3}\left(\mathbf{N} \cdot \hat{\mathbf{q}} + \frac{s}{2}\partial_q\left[(\mathbf{A}_s \cdot \mathbf{d}) \cdot \mathbf{d}\right]\right), \tag{15}$$

and

$$\partial_q\left[(\mathbf{A}_s \cdot \mathbf{d}) \cdot \mathbf{d}\right] = \left[(\mathbf{A}_s \cdot \hat{\mathbf{q}})\hat{\mathbf{I}} + \hat{\mathbf{q}} \otimes \mathbf{A}_s\right] \cdot \mathbf{d}, \quad \forall \hat{\mathbf{q}} \in \{\hat{\mathbf{x}}, \hat{\mathbf{y}}, \hat{\mathbf{z}}\}. \tag{16}$$

Hence, each term $\partial_q\left[(\mathbf{A}_s \cdot \mathbf{d}) \cdot \mathbf{d})\right]_{ij}$ can be expressed as a scalar product of a vector $(\mathbf{T}_q)_{ij}$ with $\mathbf{d}$, so that $\partial_q\left[(\mathbf{A}_s \cdot \mathbf{d}) \cdot \mathbf{d}\right] \equiv \mathbf{T}_q \cdot \mathbf{d}$, where the explicit terms of the symmetric tensors $\mathbf{T}_q$ are shown in Table S1 (Supporting Information).

Then (15) reads,

$$\partial_q \mathbf{\Delta}_{\text{pm}} = \frac{1}{3}\left(\mathbf{N} \cdot \hat{\mathbf{q}} + \frac{s}{2}\mathbf{T}_q \cdot \mathbf{d}\right). \tag{17}$$

Employing (13) and (14) the trace derivatives $\partial_q t_2$ and $\partial_q t_4$ are obtained in the Supporting Information (see, (S40) and (S41)).

Finally, it appears that to get the optimal magnetic center, we must find a solvent to a set of cubic equations in $\mathbb{R}^3$ given by $\nabla_{\mathbf{d}}(t_4 + t_2) \equiv \mathbf{0}$. Employing the results of (S40) and (S41) we can rewrite the gradient in a general matrix polynomial form as

$$\nabla_{\mathbf{d}}(t_4 + t_2) = \left[(\mathbf{A}_3 \cdot \mathbf{d}) \cdot \mathbf{d}\right] \cdot \mathbf{d} + (\mathbf{A}_2 \cdot \mathbf{d}) \cdot \mathbf{d} + \mathbf{A}_1 \cdot \mathbf{d} + \mathbf{A}_0 \equiv \mathbf{0}, \tag{18}$$

where $\mathbf{A}_0 \in \mathbb{R}$ is a 3D vector, and $\mathbf{A}_{1,2,3} \in \mathbb{R}$ are respectively $3 \times 3$, $3 \times 3 \times 3$, and $3 \times 3 \times 3 \times 3$ tensors. A diligent analysis and solution of this system goes beyond the scope of this paper and will be published elsewhere. In the following section, we consider a 1D case of axial symmetry.

## OSC$^\mathrm{m}$: the Axisymmetric Case

To demonstrate the approach, we provide a solution for the case with axial symmetry, where the axial offset $\mathbf{d} = \hat{\mathbf{z}}d$ remains the only term so that (18) degenerates into

$$\partial_z(t_4 + t_2) = a_3 z^3 + a_2 z^2 + a_1 z + a_0 \equiv 0, \tag{19}$$

with $a_0 = \frac{2}{3}\Re\left(\mathrm{Tr}\left[\mathbf{Q}^*_\mathrm{pm} \cdot (\mathbf{N} \cdot \hat{\mathbf{z}})\right]\right)$, $a_1 = \frac{2}{9}\Re\left(\mathrm{Tr}\left[\frac{3}{2}\mathbf{Q}^*_\mathrm{pm} \cdot (s\mathbf{T}_z \cdot \hat{\mathbf{z}}) + (\mathbf{N}^* \cdot \hat{\mathbf{z}}) \cdot (\mathbf{N} \cdot \hat{\mathbf{z}})\right]\right)$, $a_2 = \frac{1}{9}\Re\left(\mathrm{Tr}\left[s(\mathbf{N}^* \cdot \hat{\mathbf{z}}) \cdot (\mathbf{T}_z \cdot \hat{\mathbf{z}}) + s^*((\mathbf{A}^*_\mathrm{s} \cdot \hat{\mathbf{z}}) \cdot \hat{\mathbf{z}}) \cdot (\mathbf{N} \cdot \hat{\mathbf{z}})\right]\right)$, and $a_3 = \frac{|s|^2}{18}\Re\left(\mathrm{Tr}\left[((\mathbf{A}^*_\mathrm{s} \cdot \hat{\mathbf{z}}) \cdot \hat{\mathbf{z}}) \cdot (\mathbf{T}_z \cdot \hat{\mathbf{z}})\right]\right)$.

The number of real roots in (19) is defined by the sign of its discriminant,

$$D^{\{3\}} = -4a_3 a_1^3 + a_2^2 a_1^2 + 18 a_0 a_2 a_3 a_1 - 4 a_0 a_2^3 - 27 a_0^2 a_3^2.$$

Since in our case, the coefficients of the cubic polynomial (19) are real, then, in accordance with the complex conjugate root theorem (see e.g., p. 22 in Ref.[47]), for $D^{\{3\}} < 0$ only $d_1 \in \mathbb{R}$, i.e. is the valid solvent, while the other two roots form a complex conjugate pair ($\{d_2 = d_3^*\} \in \mathbb{C}$) and should be ignored. If $D^{\{3\}} > 0$, all three roots are real and distinct. Finally, in the degenerate case of $D^{\{3\}} = 0$, multiple roots $d_{1,2,3} \in \mathbb{R}$ are present. The solutions require a comprehensive insight into their background physics. The all-three real-root case $D^{\{3\}} > 0$ gives two possible outcomes for its ordered solvents $d_1 < d_2 < d_3$. The proper choice depends on the sign of $\partial_z^2(t_4 + t_2)|_{z=d_2}$. If $\partial_z^2(t_4 + t_2)|_{z=d_2} < 0$ then only $d_2$ becomes valid. Otherwise, for $\partial_z^2(t_4 + t_2)|_{z=d_2} > 0$ we end up with the other two valid displacements ($d_1, d_3$) for which $\partial_z^2(t_4 + t_2)$ should be negative. These cases are illustrated in the section on numerical experiments. In the case of $D^{\{3\}} < 0$ we pick the only possible real displacement of $d_1 \in \mathbb{R}$.

## Verification Case Studies

The initial successful verification of the expressions for the shifted poloidal and toroidal quadrupoles of Tables 3 and 4 was achieved numerically with an example of plane-wave scattering at a dielectric disk, (see Supporting Information). Recently, Ospanova et al.[48] presented a related set of equations for the off-center multipoles with a different normalization and limited to the poloidal terms only.

The follow-up example, shown in Fig. 2a, considers the case of a dielectric circular cone in vacuum. This axisymmetric cone is illuminated by an $x$-polarized plane wave propagating along the axis of symmetry, in the $z$-direction. A local system of coordinates, used to compute the multipoles, is shifted along the $z$-axis at a distance $d_{\text{CM}}$ from the cone CM.

We perform numerical simulations over a range of wavelengths (570 – 1000 nm) and $d_{\text{CM}}$ values, changing from $-75$ to $+75$ nm, with a proprietary surface integral equation solver.[49] Then, we compute the multipoles using the formulas in Tables 1 and 3 along with the numerically simulated electric field $\mathbf{E}$ within the particle and the contrast current defined by $\mathbf{J} = -i\omega\varepsilon_0\mathbf{E}\left(\varepsilon_{\text{r,particle}} - \varepsilon_{\text{r,bg}}\right)$, where $\varepsilon_{\text{r,particle}}$ and $\varepsilon_{\text{r,bg}}$ are the relative permittivities of the particle and the background medium, respectively. We also compute the positions of the electric ($d_{\text{e},z}$) and magnetic ($d_{\text{m},z}$) OSC using (9) and (19), respectively, and plot the results in Fig. 2b and Fig. 2c. Since the OSC positions are obtained from $d_{\text{e},z}$ and $d_{\text{m},z}$ by minimizing the electric and magnetic quadrupolar norms, we respectively plot these frequency-dispersive positions in Fig. 2b and Fig. 2c for comparison. To highlight the good match between these results, the inverted-color solid lines in (b) and (c) separately trace the values where the multipole expansions are taken at the optimal electric and magnetic centers $d_{\text{e},z}$ and $d_{\text{m},z}$. As expected, these analytically defined positions coincide very well with the minima of the quadrupolar norms.

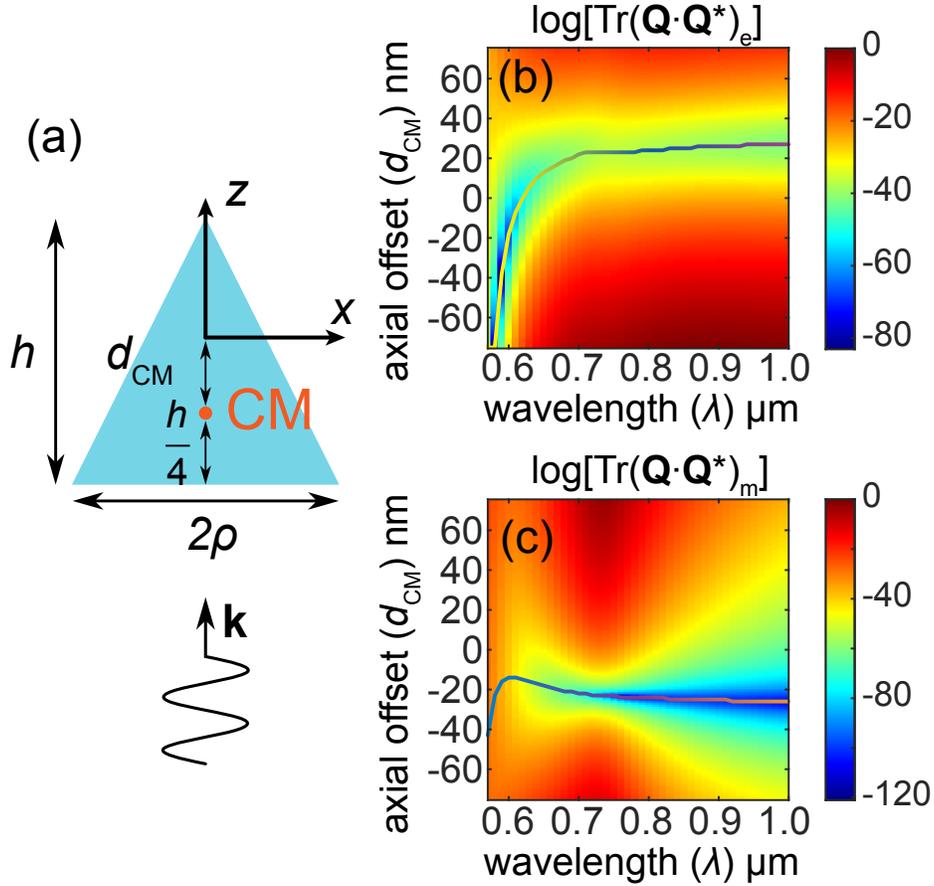

Figure 2: Optimal scattering centers of a dielectric cone. (a) A cone with permittivity $\varepsilon_r = 10$ and dimensions $h = 2\rho = 300$ nm is illuminated by an $x$-polarized plane wave. A local Cartesian coordinate system is shifted by an offset of $d_{CM}$ from the CM along the $z$-axis. (b) and (c), respectively, show the logarithmic norms of the electric and magnetic quadrupoles, $\log\left[\text{Tr}\left(\mathbf{Q}_e \cdot \mathbf{Q}_e^*\right)\right]$ and $\log\left(\text{Tr}\{\mathbf{Q}_m \cdot \mathbf{Q}_m^*\}\right)$ in dB scale. The inverted-color solid lines indicate the OSC positions obtained from $d_{e,z}$ and $d_{m,z}$.

Overall, these results illustrate an important feature that stems from the lateral asymmetry of the conical particle – *the electric and magnetic OSCs lie on opposite sides of the CM*. In general, the absolute and mutual positions of the electric and magnetic centers, their frequency and angular dispersion can be considered as important topological metrics of a given scatterer or emitter. In our case study, the electric center stabilizes at about $+25$ nm above the CM, and the magnetic center at about $-25$ nm below the CM within a broad range of $\lambda = 0.7 - 1$ $\mu$m.

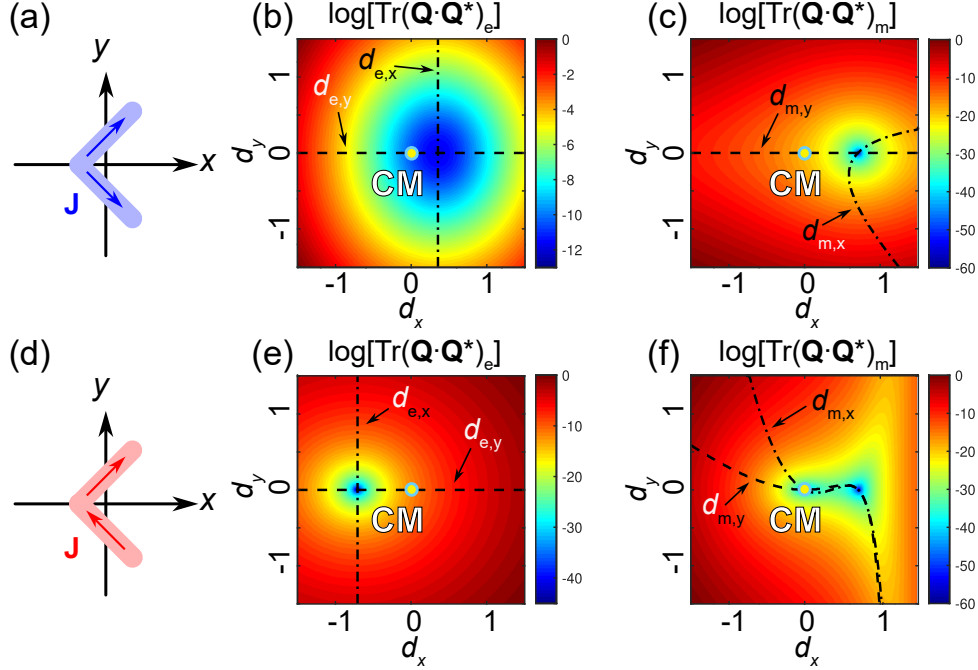

Figure 3: Optimal scattering centers of idealized current sources in a V-shaped particle. (a) and (d) show the simplified current distributions of the symmetric and anti-symmetric modes, respectively. Panels (b,e), and (c,f), respectively show the log-scale norms of the electric and magnetic quadrupole tensors versus the lateral translation of the particle (with displacements $d_x$ and $d_y$) within the $xy$-plane at $z = 0$. The dotted-dashed and dashed lines indicate the position of the OSC along the $x$- and $y$-directions, respectively, computed with (18). All calculations are done with arbitrary spatial units (a.u.).

It also becomes apparent that in a practical sense, it is *not necessary to numerically retrieve the multipoles for every value of $d_{CM}$*. Indeed, Fig. S1 in Supporting Information, confirms that it is in fact, sufficient to compute the positions of OSCs with $d_{e/m,z}$ *for any convenient, fixed choice of $d_{CM}$* (for instance, $d_{\text{CM}} = 0$, i.e., at the CM). This fixed initial reference uniquely defines the required positions of the electric and magnetic OSCs. Then, the electric and magnetic moments may be separately computed, employing the optimal positions of the multipole expansion centers $\mathbf{d} = d_{e,z}\hat{\mathbf{z}}$ and using the analytical formulas of Tables 2 and 4, with no need to perform additional, time-costly numerical multipole re-expansions of the field data.

The other example explores idealized current sources given in terms of Dirac delta distributions. We consider two current distributions that correspond to the symmetric and anti-symmetric modes of a V-shaped plasmonic particle,[50] shown in Fig. 3a and Fig. 3d, respectively.

The current distribution is given by

$$\mathbf{J}(x,y) = \mathbf{J}_+\left(\frac{1}{2\sqrt{2}}, \frac{3}{2\sqrt{2}}\right) + \mathbf{J}_+\left(-\frac{1}{2\sqrt{2}}, \frac{1}{2\sqrt{2}}\right) \\ \pm \left[\mathbf{J}_-\left(\frac{1}{2\sqrt{2}}, -\frac{3}{2\sqrt{2}}\right) + \mathbf{J}_-\left(-\frac{1}{2\sqrt{2}}, -\frac{1}{2\sqrt{2}}\right)\right], \tag{20}$$

with

$$\mathbf{J}_\pm(x', y') = (\hat{\mathbf{x}} \pm \hat{\mathbf{y}})\, \delta(x - d_x - x', y - d_y - y'), \tag{21}$$

where $\delta(x, y)$ is the Dirac delta distribution and $d_x$ and $d_y$ are spatial shifts used to displace the current distribution (20) with respect to the system of coordinates. Note that we restrict ourselves to shifts in the $xy$-plane and assume that the current distribution (20) lies at $z = 0$. The top and bottom signs in (20) correspond to the symmetric and anti-symmetric modes of the particle, respectively.

We now compute the electric and magnetic quadrupolar norms by inserting (20) into the formulas of Table 1 and plot the results in Fig. 3b and Fig. 3c, for the symmetric mode, and in Fig. 3e and Fig. 3d, for the anti-symmetric mode, respectively. In these plots, the black dashed, and dotted-dashed lines correspond to the minima (where $d_{\mathrm{e/m},x/y} \approx 0$) of the distances to the electric and magnetic OSC along the $x$- and $y$-directions computed using (18). As expected, the $d_{\mathrm{e/m},x}$ lines intersect the $d_{\mathrm{e/m},y}$ lines at the positions of the minima of the quadrupolar norms.

In the plots of Fig. 3, we have also indicated the position of the CM. Since the CM offset from the external corner is given by $\frac{l^2 + lw - w^2}{\sqrt{2}(2l-w)}$, where $l$ is the length of the unrounded leg, measured from the external corner, and $w$ is its width. For a very small width (our case), the offset from the corner along the axis of symmetry degenerates into $\frac{1}{2\sqrt{2}}l$. In our case, $l = 2$ a.u. and the CM is at the origin of the system of coordinates. Comparing the positions of the OSC$^\mathrm{e}$ and OSC$^\mathrm{m}$, we see that they do not coincide with each other and that they are both offset from the CM, as was the case in Fig. 2, at the exception of one the OSC$^{\text{m}}$ in Fig. 3f which coincides with the CM. Interestingly, the anti-symmetric mode exhibits two magnetic OSCs, as can be seen in Fig. 3f. Finally, note that for the symmetric and anti-symmetric cases, there are three solutions for $d_{\text{m},y}$ and $d_{\text{m},x}$, respectively. One of these solutions is purely real (and is plotted in Fig. 3c and in Fig. 3f) and the two others are complex conjugated pairs; this result is consistent with the fundamental theorem of algebra and the cubic nature of Eq. (19), as discussed in the previous section.

## Summary and Discussions

Optimizing the multipole descriptions by using ultimate expansion centers has been limited to the quasistatic approximations. Expanding this critical problem beyond electro- and magnetostatics can enable new advanced theoretical concepts and efficient numerical schemes. Here, we use a truncated set of poloidal Cartesian multipoles[10,42–45] to demonstrate the concept and derive the ultimate positions for the distinct reference points for the electric and magnetic multipole expansions that minimize the norm of matching poloidal quadrupoles. While this work considers a truncated poloidal dipole-quadrupole model, further expansions to rigorous spherical multipoles or high-order poloidal Cartesian moments are possible and will be discussed elsewhere. The shifted toroidal multipoles (electric and magnetic) are also derived and presented separately in Supporting Information.

The above results provide a base for a comprehensive understanding of center-splitting required to get the best multipolar descriptor. Once the position of $O'_e$ is optimized, it would become beneficial *to keep $O'_e$ for the electric multipole expansion*, while *seeking a new, separate expansion center $O'_e$ for the magnetic multipoles*.

Figure 4 illustrates the splitting of centers, $O'_e$ and $O'_m$. The immediate consequence of

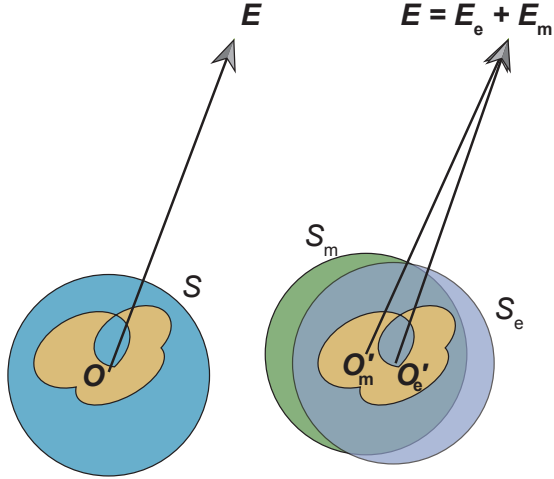

Figure 4: Splitting the positions of multipole scattering centers. The left panel depicts the non-optimized position of a joint expansion center $O$. The right panel shows two separate positions of optimized expansion centers ($O'_\text{e}$ and $O'_\text{m}$) along with two different basis surfaces ($S_\text{e}$ and $S_\text{m}$).

this splitting appears in (3) as

$$\mathbf{E} = \kappa_\text{e} \left[ \hat{\mathbf{r}}_\text{e} \times (\mathbf{p} - ik\,\mathbf{Q}_\text{e} \cdot \hat{\mathbf{r}}_\text{e}) \right] \times \hat{\mathbf{r}}_\text{e}$$

$$+ \kappa_\text{m} \left( \mathbf{m} - ik\,\mathbf{Q}_\text{m} \cdot \hat{\mathbf{r}}_\text{m} \right) \times \hat{\mathbf{r}}_\text{m},$$

with

$$\kappa_\text{e} = \frac{k^2 \mathrm{e}^{ikr_\text{e}}}{4\pi\varepsilon_0 r_\text{e}},\ \kappa_\text{m} = \frac{k^2 \mathrm{e}^{ikr_\text{m}}}{4\pi z_0 r_\text{m}}.$$

As shown in the right panel of Fig. 4, the contribution to the electric field from electric multipoles ($\mathbf{E}_\text{e}$) is computed with the expansion taken relative to optimized center $O'_\text{e}$ or a new basis surface $S_\text{e}$. Conversely, the contribution $\mathbf{E}_\text{m}$ from magnetic multipoles ($\mathbf{p}$, $\mathbf{Q}_\text{e}$) is computed with their expansion taken relative to optimized center $O'_\text{m}$, i.e., using a different basis surface $S_\text{m}$. Apparently, $\mathbf{r}_\text{m} = \mathbf{r}_\text{e} + \mathbf{d}_\text{m}$.

We demonstrate the utility of our approach using a couple of representative case studies. Our numerical experiments also confirm that the optimal reference centers for the most efficient multipole models of scatterers and emitters should be obtained separately for the electric and magnetic multipole expansions. In the first case, the optimal centers calcu-lated with our analytical formulae perfectly match the positions obtained by minimizing the quadrupolar norms numerically. In the other case, we study an in-plane current model of a plasmonic V-shape antenna with the positions of the in-plane centers also matching the expected locations. The positions of the optimal centers are frequency dispersive and generally are not collocal with the center of mass, which exhibits neither frequency nor angular dependence. The optimal electric and magnetic centers present a crucial combined property of the scatterer/emitter material composition, shape, frequency or angular dispersion, and ambient. We also demonstrate in passing that the number of the optimal magnetic centers connected to the multiplicity of the roots can be linked to the symmetry and topology of the modes excited within a given scatterer. These results could already be used for fundamental studies in quantum and topological photonics. The significant reduction of computational redundancy will benefit the new efficient differentiable solvers, including the generalized T-matrix and Multiple Multipole schemes, which are critical for the machine-learning-driven inverse-design frameworks in photonics and optoelectronics.

# Acknowledgement

A.V.K. acknowledges support from the Office of Naval Research under Award N00014-20-1-2199 and the Air Force Office of Scientific Research under Award FA9550-21-1-0299. A.V.K. wants to thank Prof. I. Shadrivov, TMOS (ANU, Canberra, Australia), and Prof. O. J. Martin, EPFL (Lausanne, Switzerland) for their kind support of his sabbatical visits. This support was critical for the progress of this effort. K.A. acknowledges funding from the Swiss National Science Foundation (project PZ00P2_193221). D.S. acknowledges support from the Australian Research Council (FT230100058).

# Supporting Information Available

Sections Electric and Magnetic Multipoles of the Supporting information contain the derivation details and explicit formulations of the tensor forms for the unshifted and shifted (off-center) electric and magnetic moments, which are then summarized in Tables 1–4 of the main text. Section *Ad Hoc* Lemma deals with a problem-specific approach to inverting the system matrix of a given symmetry and structure. Finally, the Explicit Forms section shows the explicit matrix forms of auxiliary tensors $\mathbf{A}_s$, $\mathbf{T}_q$, and trace derivatives $\partial_q t_2$ and $\partial_q t_4$.

# SUPPORTING INFORMATION: THE ART OF FINDING THE OPTIMAL SCATTERING CENTER(S)

Alexander V. Kildishev,*,†,‡ Karim Achouri,¶ and Daria Smirnova§

†Elmore Family School of Electrical and Computer Engineering and Birck Nanotechnology Center, Purdue University, 1205 W State St, West Lafayette, IN 47907, USA

‡On sabbatical leave: ARC Centre of Excellence for Transformative Meta-Optical Systems (TMOS), Research School of Physics, Australian National University, Canberra, ACT 2601, Australia, and the Nanophotonics and Metrology Laboratory, Institute of Electrical and Microengineering, École Polytechnique Fédérale de Lausanne, Route Cantonale, 1015 Lausanne, Switzerland

¶Nanophotonics and Metrology Laboratory, Institute of Electrical and Microengineering, École Polytechnique Fédérale de Lausanne, Route Cantonale, 1015 Lausanne, Switzerland.

§ARC Centre of Excellence for Transformative Meta-Optical Systems (TMOS), Research School of Physics, Australian National University, Canberra, ACT 2601, Australia

E-mail: kildishev@purdue.edu

Phone: +1 765 586 7435

# Electric Multipoles

## Unshifted Electric Dipole

### Unshifted Poloidal Electric Dipole

As shown in Introduction, a classical series expansion for a small argument gives an LWA to the spherical Bessel function. Then, the poloidal part of an unshifted electric dipole is[1]

$$\mathbf{p}_0 = \frac{1}{s} \int dV \mathbf{J}, \tag{S1}$$

and its axial vector components are defined as

$$\mathbf{p}_0 \triangleq \hat{\mathbf{x}}\, p_{0x} + \hat{\mathbf{y}}\, p_{0y} + \hat{\mathbf{z}}\, p_{0z}. \tag{S2}$$

### Unshifted Toroidal Electric Dipole

The toroidal part is split into two integrals given by

$$\mathbf{p}_{\text{te1}} = a \int dV (\mathbf{r} \times (\mathbf{r} \times \mathbf{J})), \tag{S3}$$

and

$$\mathbf{p}_{\text{te2}} = -a \int dV (\mathbf{r} \cdot \mathbf{r}) \mathbf{J}, \tag{S4}$$

with

$$a = -\frac{k^2}{10s}. \tag{S5}$$

### Unshifted Toroidal Electric Dipole: Alternative Definition

Alternatively, we may get another definition that is used later on. First, an auxiliary tensor $\mathbf{T}_e$ is defined as follows

$$\mathbf{T}_e = a \int dV \left[ \mathbf{J} \otimes (\mathbf{r} \otimes \mathbf{r}) \right]. \tag{S6}$$

Then, the tensor $\mathbf{T}_e$ is used to get the components of the toroidal electric dipole

$$\mathbf{p}_{te} = \mathbf{p}_{t1} + \mathbf{p}_{t2}, \tag{S7}$$

with $\mathbf{p}_{t1}$ that matches integral (8)

$$\mathbf{p}_{t1} = \mathbf{p}_{t2} + \sum_{\hat{\mathbf{q}}}^{\{\hat{\mathbf{x}},\hat{\mathbf{y}},\hat{\mathbf{z}}\}} \hat{\mathbf{q}} \cdot (\hat{\mathbf{q}} \cdot \mathbf{T}_e), \tag{S8}$$

and $\mathbf{p}_{t2}$ that matches integral (9)

$$\mathbf{p}_{t2} = - \sum_{\hat{\mathbf{q}}}^{\{\hat{\mathbf{x}},\hat{\mathbf{y}},\hat{\mathbf{z}}\}} (\mathbf{T}_e \cdot \hat{\mathbf{q}}) \cdot \hat{\mathbf{q}}. \tag{S9}$$

While the projections of the tensor $\mathbf{T}_e$ to a lower dimension in (S8) and (S9) give the values of $\mathbf{p}_{t1,2}$, the tensor is directly used in the definition of the shifted toroidal multipoles, shown below in (S22), (S23), (S39), and summarized in Table 4 of the main text. Upfront computing of $\mathbf{T}_e$ eliminates the need to recompute the volume integral for all these toroidal multipoles, provided that $\mathbf{T}_e$ is already calculated.

### Unshifted High-order Toroidal Electric Dipole

An unshifted high-order toroidal correction to the electric dipole can be written as[1]

$$\mathbf{p}_{t3} = \frac{ak^2}{28} \int dV \left[ 3\mathbf{J} - 2(\hat{\mathbf{r}} \otimes \hat{\mathbf{r}}) \cdot \mathbf{J} \right] r^4. \tag{S10}$$

Note that we use a slightly different normalization as in Ref.[1]

## Shifted Toroidal Electric Dipole

Once the origin is shifted, we end up with a shifted expansion, and the shifted toroidal dipolar terms are

$$\begin{aligned}\mathbf{p}'_{t1} &= \mathbf{p}_{t1} + 2a\left(\mathbf{d}\times(\mathbf{m}_0 + \frac{s}{2}\mathbf{d}\times\mathbf{p}_0) + s\left[\text{Tr}(\mathbf{Q}_{e1})\mathbf{d} - \mathbf{d}\cdot\mathbf{Q}_{e1})\right]\right)\\ &= \mathbf{p}_{t1} + 2a\left(\mathbf{d}\times\mathbf{m}'_0 + s\left[\text{Tr}(\mathbf{Q}_{e1})\mathbf{d} - \mathbf{d}\cdot\mathbf{Q}_{e1}\right]\right),\end{aligned} \quad (S11)$$

$$\mathbf{p}'_{t2} = \mathbf{p}_{t2} - as\left(d^2\,\mathbf{p}_0 + 4\mathbf{d}\cdot\mathbf{Q}_{e1}\right), \quad (S12)$$

where the unshifted poloidal quadrupole $\mathbf{Q}_{e1}$ and the shifted poloidal magnetic dipole $\mathbf{m}'_0 = \mathbf{m}_0 + \frac{s}{2}\mathbf{d}\times\mathbf{p}_0$ are defined in Tables 1 and 2.

For an arbitrary offset, the components of the shifted electric dipole are defined by the contributions from the translation-invariant dipole $\mathbf{p}_0$, *unshifted poloidal quadrupole* $\mathbf{Q}_{e1}$, and *unshifted poloidal magnetic dipole* $\mathbf{m}_0$. These contributions lead to

$$\begin{aligned}\mathbf{p}' &= \mathbf{p} + 2a\left[\mathbf{d}\times(\mathbf{m}_0 + \frac{s}{2}\mathbf{d}\times\mathbf{p}_0) - s\left(\frac{1}{2}d^2\mathbf{p}_0 + 3\mathbf{d}\cdot\mathbf{Q}_{e1} - \text{Tr}(\mathbf{Q}_{e1})\mathbf{d}\right)\right]\\ &= \mathbf{p} + 2a\left[\mathbf{d}\times\mathbf{m}'_0 - s\left(\frac{1}{2}d^2\mathbf{p}_0 + 3\mathbf{d}\cdot\mathbf{Q}_{e1} - \text{Tr}(\mathbf{Q}_{e1})\mathbf{d}\right)\right].\end{aligned} \quad (S13)$$

We recap that only the poloidal dipolar term $\mathbf{p}_0$ is unaffected by any translation (i.e., $\mathbf{p}'_0 \equiv \mathbf{p}_0$), while the translation-dependent toroidal terms $\mathbf{p}'_{1,2}$ receive contributions from $\mathbf{p}_0$, $\mathbf{m}_0$, and $\mathbf{Q}_{e1}$.

### Shifted High-order Toroidal Electric Dipole

The shifted high-order toroidal correction to the electric dipole yields a heavy formula that includes unshifted components of all kinds,

$$\begin{aligned}\mathbf{p}'_{t3} = \mathbf{p}_{t3} &- \left(\frac{ak^2 s}{14}\mathbf{d}\cdot\mathbf{d}\left(\frac{1}{2}\mathbf{d}\cdot\mathbf{d}\mathbf{p}_0 - \mathbf{d}\times(\mathbf{d}\times\mathbf{p}_0) + 6\mathbf{d}\cdot\mathbf{Q}_{e1} - 2\text{Tr}[\mathbf{Q}_{e1}]\mathbf{d}\right)\right.\\ &-\frac{1}{14}\left(ak^2\right)\mathbf{d}\times\left(\mathbf{d}\cdot(12\mathbf{Q}_{m1} + 2\mathbf{d}\otimes\mathbf{m}_0) + \mathbf{d}\times\left(4s\mathbf{d}\cdot\mathbf{Q}_{e1} - \frac{\mathbf{p}_{t2}}{a}\right) - \frac{2}{as}\mathbf{m}_t\right)\\ &-\frac{1}{14a}\mathbf{d}\cdot\mathbf{d}\left(ak^2\right)(\mathbf{p}_{t1}+\mathbf{p}_{t2}) + \frac{1}{7}\left(2k^2\right)\mathbf{d}\cdot\left((\mathbf{T}_e)^T\cdot\mathbf{d} - \frac{21}{10}\left(\frac{\mathbf{Q}_{e2}}{2}+\mathbf{Q}_{e3}\right)\right)\\ &-\frac{k^2}{7}\mathbf{d}\left(\mathbf{d}\cdot(\mathbf{p}_{t1}-\mathbf{p}_{t2}) - \frac{21}{25}\text{Tr}[\mathbf{Q}_{e3}]\right)\bigg).\end{aligned}$$

## Unshifted Electric Quadrupole

LWA of an electric quadrupole at a non-shifted origin includes its poloidal and toroidal parts, hence we define the *unshifted electric quadrupole* as $\mathbf{Q}_e = \mathbf{Q}_{pe} + \mathbf{Q}_{te}$.

### Unshifted Poloidal and Toroidal Electric Quadrupole

The poloidal electric quadrupole

$$\mathbf{Q}_{pe} = (\mathbf{Q}_{e1})^S - \frac{2}{3}\hat{\mathbf{I}}\text{Tr}\left[\mathbf{Q}_{e1}\right] \tag{S14}$$

is defined using the single integral

$$\mathbf{Q}_{e1} = \frac{1}{2s}\int dV(\mathbf{r}\otimes\mathbf{J}). \tag{S15}$$

In contrast, the toroidal part

$$\mathbf{Q}_{te} = \mathbf{Q}_{e2} + \frac{1}{2}\hat{\mathbf{I}}\text{Tr}\left(\mathbf{Q}_{e2}\right) + (\mathbf{Q}_{e3})^S \tag{S16}$$

is defined through

$$\mathbf{Q}_{e2} = \frac{10}{21}a \int dV (\mathbf{r} \otimes \mathbf{r})(\mathbf{r} \cdot \mathbf{J}), \tag{S17}$$

$$\mathbf{Q}_{e3} = -\frac{25}{42}a \int dV (\mathbf{r} \cdot \mathbf{r})(\mathbf{r} \otimes \mathbf{J}). \tag{S18}$$

A straightforward analysis of the sums of poloidal and toroidal parts shows that both parts are *individually traceless*, i.e.,

$$\mathrm{Tr}\,(\mathbf{Q}_{\mathrm{pe}}) \equiv 0; \quad \mathrm{Tr}\,(\mathbf{Q}_{\mathrm{te}}) \equiv 0, \tag{S19}$$

thus confirming that entire electric quadrupole $\mathbf{Q}_{\mathrm{e}} = \mathbf{Q}_{\mathrm{pe}} + \mathbf{Q}_{\mathrm{te}}$ is traceless.

## Shifted Electric Quadrupole

We also define the *shifted electric quadrupole*, splitting it into the shifted poloidal ($\mathbf{Q}'_{\mathrm{pe}} = (\mathbf{Q}'_{\mathrm{e1}})^{\mathrm{S}} - \frac{2}{3}\hat{\mathbf{I}}\mathrm{Tr}\,[\mathbf{Q}'_{\mathrm{e1}}]$) and toroidal ($\mathbf{Q}'_{\mathrm{te}} = \mathbf{Q}'_{\mathrm{e2}} + \frac{1}{2}\hat{\mathbf{I}}\mathrm{Tr}\,[\mathbf{Q}'_{\mathrm{e2}}] + (\mathbf{Q}'_{\mathrm{e3}})^{\mathrm{S}}$) parts.

### Shifted Poloidal Electric Quadrupole

The poloidal part yields

$$\mathbf{Q}'_{\mathrm{e1}} = \mathbf{Q}_{\mathrm{e1}} + \frac{1}{2}\mathbf{d} \otimes \mathbf{p}_0. \tag{S20}$$

Then, inserting the above into $\mathbf{Q}'_{\mathrm{pe}} = (\mathbf{Q}'_{\mathrm{e1}})^{\mathrm{S}} - \frac{2}{3}\hat{\mathbf{I}}\mathrm{Tr}\,[\mathbf{Q}'_{\mathrm{e1}}]$ gives the shifted poloidal part

$$\mathbf{Q}'_{\mathrm{pe}} = \mathbf{Q}_{\mathrm{pe}} + \frac{1}{2}(\mathbf{d} \otimes \mathbf{p}_0)^{\mathrm{S}} - \frac{1}{3}\hat{\mathbf{I}}(\mathbf{d} \cdot \mathbf{p}_0). \tag{S21}$$

Only the translation-invariant dipole $\mathbf{p}_0$ contributes to the poloidal part of shifted quadrupoles.

### Shifted Toroidal Electric Quadrupole

The shifted toroidal part $\mathbf{Q}'_{\text{te}} = \mathbf{Q}'_{\text{e2}} + \frac{1}{2}\hat{\mathbf{I}}\text{Tr}\left[\mathbf{Q}'_{\text{e2}}\right] + \mathbf{Q}'_{\text{e3}}$ appears to be more intricate. It may be defined using

$$\begin{aligned}\mathbf{Q}'_{\text{e2}} = \mathbf{Q}_{\text{e2}} &+ \frac{10}{21}\mathbf{d}\cdot\mathbf{T}_{\text{e}} \\ &+ \frac{20}{21}as\left[\left(\frac{1}{2as}\mathbf{d}\otimes(\mathbf{p}_{\text{t1}} - \mathbf{p}_{\text{t2}}) + \mathbf{Q}_{\text{e1}}\cdot(\mathbf{d}\otimes\mathbf{d})\right)^{\text{S}} + \text{Tr}\left(\mathbf{Q}_{\text{e1}}\right)(\mathbf{d}\otimes\mathbf{d})\right],\end{aligned} \quad (S22)$$

and

$$\begin{aligned}\mathbf{Q}'_{\text{e3}} = &\left(\mathbf{Q}_{\text{e3}} - \frac{25}{21}\mathbf{T}_{\text{e}}\cdot\mathbf{d}\right)^{\text{S}} \\ &- \frac{25}{21}as\left(d^2\mathbf{Q}'_{\text{e1}} - \frac{1}{2as}\mathbf{d}\otimes\mathbf{p}_{\text{t2}} + 2(\mathbf{d}\otimes\mathbf{d})\cdot\mathbf{Q}_{\text{e1}}\right)^{\text{S}}.\end{aligned} \quad (S23)$$

The poloidal ($\mathbf{Q}'_{\text{pe}}$) and toroidal ($\mathbf{Q}'_{\text{te}}$) parts are also individually traceless, i.e.,

$$\text{Tr}\left(\mathbf{Q}'_{\text{pe}}\right) \equiv 0; \quad \text{Tr}\left(\mathbf{Q}'_{\text{te}}\right) \equiv 0. \quad (S24)$$

# Magnetic Multipoles

## Unshifted Magnetic Dipole

### Unshifted Poloidal Magnetic Dipole

A similar approach to the components of the unshifted magnetic dipole gives

$$\mathbf{m}_0 = \frac{1}{2}\int \text{d}V(\mathbf{r}\times\mathbf{J}). \quad (S25)$$

### Unshifted Toroidal Magnetic Dipole

To match (3), we define the normalized unshifted toroidal magnetic dipole as

$$\mathbf{m}_\mathrm{t} = -\frac{as}{2} \int \mathrm{d}V (\mathbf{r} \cdot \mathbf{r})(\mathbf{r} \times \mathbf{J}). \tag{S26}$$

## Shifted Magnetic Dipole

### Shifted Poloidal Magnetic Dipole

Note that the shifted components of the poloidal magnetic dipole are affected solely by the translation-invariant part of electric dipole $\mathbf{p}_0$, leading to

$$\mathbf{m}_0' = \mathbf{m}_0 + \frac{s}{2}\mathbf{d} \times \mathbf{p}_0. \tag{S27}$$

### Shifted Toroidal Magnetic Dipole

Taking into account (S27), the shifted toroidal magnetic dipole yields

$$\mathbf{m}_\mathrm{t}' = \mathbf{m}_\mathrm{t} - as\left[d^2\mathbf{m}_0' - \frac{1}{2a}\mathbf{d} \times (\mathbf{p}_{\mathrm{t}2} - 4as\mathbf{d} \cdot \mathbf{Q}_{\mathrm{e}1}) + 6\mathbf{d} \cdot \mathbf{Q}_{\mathrm{m}1}\right]. \tag{S28}$$

## Unshifted Magnetic Quadrupole

LWA of a non-shifted magnetic quadrupole contains poloidal and toroidal parts, so we define the *unshifted magnetic quadrupole* as $\mathbf{Q}_\mathrm{m} = \mathbf{Q}_\mathrm{pm} + \mathbf{Q}_\mathrm{tm}$.

### Unshifted Poloidal Magnetic Quadrupole

Employing a general vector product notation, we may define a partial kernel of the magnetic quadrupole as

$$\mathbf{Q}_{\mathrm{m}1} = \frac{1}{6}\int \mathrm{d}V\, [\mathbf{r} \otimes (\mathbf{r} \times \mathbf{J})] \equiv \frac{1}{6}\int \mathrm{d}V\, [(\mathbf{r} \otimes \mathbf{r}) \times \mathbf{J}] \tag{S29}$$

to obtain the total poloidal magnetic quadrupole as a sum of two transposed tensors

$$\mathbf{Q}_{\mathrm{pm}} = (\mathbf{Q}_{\mathrm{m1}})^{\mathrm{S}}. \tag{S30}$$

Alternatively, employing a general vector product notation, we may write

$$\mathbf{Q}_{\mathrm{pm}} = \frac{1}{6}\left[\int \mathrm{d}V(\mathbf{r}\otimes\mathbf{r}\times\mathbf{J})\right]^{\mathrm{S}}. \tag{S31}$$

### Unshifted Toroidal Magnetic Quadrupole

Employing a general vector product notation, we may define a partial kernel of the magnetic quadrupole as

$$\mathbf{Q}_{\mathrm{m2}} = -\frac{as}{84}\int \mathrm{d}V\left(r^2\mathbf{r}\otimes\mathbf{r}\times\mathbf{J}\right) \tag{S32}$$

to obtain the total poloidal magnetic quadrupole as a sum of two transposed tensors

$$\mathbf{Q}_{\mathrm{tm}} = (\mathbf{Q}_{\mathrm{m2}})^{\mathrm{S}}. \tag{S33}$$

The poloidal and toroidal magnetic quadrupoles are traceless, i.e., $\mathrm{Tr}\,(\mathbf{Q}_{\mathrm{pm}}) \equiv 0$; $\mathrm{Tr}\,(\mathbf{Q}_{\mathrm{tm}}) \equiv 0$.

## Shifted Magnetic Quadrupole

Similarly, *a shifted magnetic quadrupole* is defined as $\mathbf{Q}'_{\mathrm{m}} = \mathbf{Q}'_{\mathrm{pm}} + \mathbf{Q}'_{\mathrm{tm}}$.

### Shifted Poloidal Magnetic Quadrupole

We employ some immediate substitutions for the shifted magnetic quadrupole, observing that it receives contributions from the poloidal electric dipole $\mathbf{p}_0$, electric quadrupole $\mathbf{Q}_{\mathrm{e1}}$, and magnetic dipole moment $\mathbf{m}_0$. We illustrate that by employing tensor-vector cross products

in the integral as

$$\mathbf{Q}'_{m1} = \mathbf{Q}_{m1} + \frac{1}{3}\left(\mathbf{d} \otimes (\mathbf{m}_0 - \frac{s}{2}\mathbf{d} \times \mathbf{p}_0) - s\,\mathbf{Q}_{e1} \times \mathbf{d}\right). \tag{S34}$$

Since $\mathbf{Q}'_{\text{pm}}$, $\mathbf{Q}_{\text{pm}}$, and $((\mathbf{d} \otimes \mathbf{d}) \times \mathbf{p}_0)^{\text{S}}$ are traceless, then

$$\text{Tr}\left[(\mathbf{d} \otimes \mathbf{m}_0 - s\,\mathbf{Q}_{e1} \times \mathbf{d})^{\text{S}}\right] \equiv 0. \tag{S35}$$

Finally, we employ (S34) to get the shifted poloidal magnetic quadrupole

$$\mathbf{Q}'_{\text{pm}} = (\mathbf{Q}'_{m1})^{\text{S}}. \tag{S36}$$

**Shifted Toroidal Magnetic Quadrupole**

First of all, we introduce an *ad hoc* mapping $\text{M} : \mathbf{w} \otimes (\mathbf{v} \otimes \mathbf{v}) \to (\mathbf{v} \otimes \mathbf{v}) \otimes \mathbf{w}$ that we cast in the short-cut notation

$$[\mathbf{w} \otimes (\mathbf{v} \otimes \mathbf{v})]^{\text{M}} \equiv (\mathbf{v} \otimes \mathbf{v}) \otimes \mathbf{w}. \tag{S37}$$

Then, substituting $\mathbf{T}_e$ in (S37) yields

$$[\mathbf{T}_e]^{\text{M}} = a \int dV \left[(\mathbf{r} \otimes \mathbf{r}) \otimes \mathbf{J}\right]. \tag{S38}$$

Apparently, the direct mapping $\text{M} : \mathbf{w} \otimes (\mathbf{v} \otimes \mathbf{v}) \to (\mathbf{v} \otimes \mathbf{v}) \otimes \mathbf{w}$ eliminates the need to recompute the volume integral, provided that the tensor $\mathbf{T}_e$ is already available.

Finally, we end up with a heavy formula for a shifted toroidal magnetic quadrupole

$\mathbf{Q}'_{\text{tm}} = (\mathbf{Q}'_{\text{m2}})^{\text{S}}$, where

$$\begin{aligned}\mathbf{Q}'_{\text{m2}} = \mathbf{Q}_{\text{m2}} &+ \frac{1}{21}\mathbf{d}\cdot(\mathbf{T}_{\text{m}} + \frac{s}{2}[\mathbf{T}_{\text{e}}]^{\text{M}}\times\mathbf{d}) \\ &- \frac{as}{7}\mathbf{d}\otimes\left(\mathbf{d}\cdot\mathbf{Q}_{\text{m1}} + \frac{s}{3}\mathbf{d}\times(\mathbf{d}\cdot\mathbf{Q}_{\text{e1}})\right) \\ &- \frac{asd^2}{14}\mathbf{Q}'_{\text{pm}} + \frac{1}{42}\mathbf{d}\otimes\left(\frac{s}{2}\mathbf{d}\times\mathbf{p}_{\text{t2}} + \mathbf{m}_{\text{t}}\right) \\ &- \frac{s}{50}\mathbf{Q}_{\text{e3}}\times\mathbf{d}.\end{aligned} \quad (S39)$$

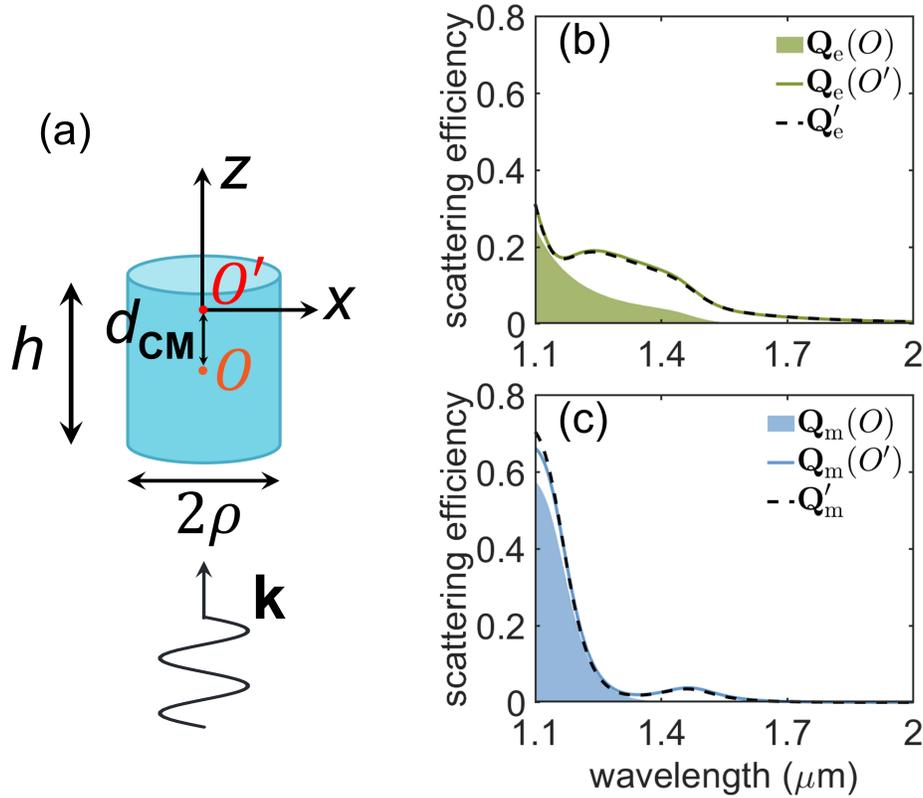

Figure S1: Verification of the LWA formulas for the order $n = 2$ with an example of a dielectric disk. The simulated disk is of the height $h = 300\,\text{nm}$, radius $\rho = 250\,\text{nm}$ and the dielectric constant $\varepsilon_{\text{r}} = 10$. The wave vector of the incident plane wave is aligned with the rotational axis $z$. Contributions to the scattering efficiency from the Cartesian electric ($\mathbf{Q}_{\text{e}}$ in (b)) and magnetic ($\mathbf{Q}_{\text{m}}$ in (c)) quadrupoles are calculated numerically – first, using the CM $O$ as the reference point for the multipole expansion (filled green and blue curves in (b) and (c), respectively). Then, the scattering efficiencies (solid green and blue lines in (b) and (c)) are also obtained numerically, employing an off-center point $O'$, shifted from the CM by 50 nm along $z$. In the LWA limit, the scattering curves for the off-center point $O'$ are well approximated by the results from the formulas of Tables 3, 4 (dashed black lines).

We verify the expressions summarized in Tables 3 and 4 for the shifted multipoles with an example of plane-wave scattering at a dielectric disk, depicted in Fig. S1a. The total scattering efficiency, and its exact expansion in spherical multipoles, are well approximated by the lowest-order Cartesian multipoles ($n = 1, 2$) at wavelengths $\lambda > 0.85$ $\mu$m. The verification results, shown in Fig. S1bc include contributions from the the poloidal and toroidal parts of the Cartesian quadrupole moments.

## *Ad Hoc* Lemma

Suppose that a nonsingular $3 \times 3$ real symmetric matrix $\mathbf{A}$ is given as a sum $\mathbf{A} = a\left(\hat{\mathbf{I}} + \mathbf{S}\right)$ of identity matrix $\hat{\mathbf{I}}$ and a real symmetric matrix $\mathbf{S}$. The sum is scaled by the scalar factor $a \in \mathbb{R}$.

Our *ad hoc* matrix lemma states that, (i) *the determinant of* $\mathbf{A}$ *is* $|\mathbf{A}| = a^3(\text{Tr}[\text{Adj}(\mathbf{S})] + |\mathbf{S}| + \text{Tr}(\mathbf{S}) + 1)$ and, (ii) *the adjugate of* $\mathbf{A}$ *is* $\text{Adj}(\mathbf{A}) = a^2(\text{Adj}(\mathbf{S}) + \hat{\mathbf{I}}(\text{Tr}(\mathbf{S}) + 1) - \mathbf{S})$.

### Ad Hoc Lemma: Singular Case

We separately consider a singular case of inverting a symmetric matrix $\mathbf{A} = a\left(\hat{\mathbf{I}} + \mathbf{S}\right)$ with $\mathbf{S} = \frac{1}{3a}\Re(\mathbf{v} \otimes \mathbf{v}^*)$ and $a = \mathbf{v} \cdot \mathbf{v}^*$, ($\forall\, \mathbf{v} \in \mathbb{C}^3$). The determinant of such a special matrix $\mathbf{S}$ is a sum of three vanishing terms, $\frac{1}{4}\left(v_2^{*2}v_1^2 + v_1^{*2}v_2^2\right)(|v_3|^2 - |v_3|^2)$, $(|v_1|\,|v_2|\,|v_3|)^2 - (|v_1|\,|v_2|\,|v_3|)^2$, and $\frac{1}{4}(|v_1|^2 - |v_1|^2)(v_3^{*2}v_2^2 + v_2^{*2}v_3^2)$, then, $|\mathbf{S}| = 0$, and $\mathbf{S}$ is *a singular matrix*. We also note that $\text{Tr}(\mathbf{S}) = \frac{1}{3}$.

For such a matrix, the above lemma degenerates into three simple statements,

(i) the determinant of $\mathbf{A}$ is
$$|\mathbf{A}| = \frac{1}{3}a^3(3\text{Tr}[\text{Adj}(\mathbf{S})] + 4),$$

(ii) the adjugate of $\mathbf{A}$ is
$$\text{Adj}(\mathbf{A}) = \frac{1}{3}a^2\left[3(\text{Adj}(\mathbf{S}) - \mathbf{S}) + 4\hat{\mathbf{I}}\right],$$

and, (iii) the inverse of $\mathbf{A}$ is

$$\mathbf{A}^{-1} = \frac{3(\text{Adj}(\mathbf{S}) - \mathbf{S}) + 4\hat{\mathbf{I}}}{a(3\text{Tr}[\text{Adj}(\mathbf{S})] + 4)}.$$

Several observations connected to this case could be instrumental for physical applications. Firstly, we note that with $\mathbf{t} \to \frac{1}{6a}\mathbf{v} \times \mathbf{v}^*$, we have that

$$\text{Adj}(\mathbf{S}) = \frac{(\mathbf{v} \times \mathbf{v}^*) \otimes (\mathbf{v}^* \times \mathbf{v})}{(6\,\mathbf{v} \cdot \mathbf{v}^*)^2} = \mathbf{t} \otimes \mathbf{t}^*,$$

then, we immediately arrive at

$$\text{Tr}(\text{Adj}(\mathbf{S})) = \frac{(\mathbf{v} \times \mathbf{v}^*) \cdot (\mathbf{v}^* \times \mathbf{v})}{(6\,\mathbf{v} \cdot \mathbf{v}^*)^2} = \mathbf{t} \cdot \mathbf{t}^*,$$

and finally, we get

$$\mathbf{A}^{-1} = \frac{3\Re(\mathbf{t} \otimes \mathbf{t}^*) - \frac{1}{a}\Re(\mathbf{v} \otimes \mathbf{v}^*) + 4\hat{\mathbf{I}}}{a(3\mathbf{t} \cdot \mathbf{t}^* + 4)}.$$

The topological significance of the vector $\mathbf{t}$ and tensor $\text{Adj}(\mathbf{S})$ for radiating point sources is discussed in the main text. More general cases for the higher-rank symmetrized outer products of non-conjugate vectors with dimensions higher than three are also available; they go beyond the scope of the current effort and will be published elsewhere.

# Explicit Forms

## Explicit Matrix Form of Tensor $\mathbf{A}_s$

We define the symmetric $3\times 3\times 3\times 3$ tensor $\mathbf{A}_s$ by the nested array

$$\mathbf{A}_s = \begin{pmatrix} \begin{pmatrix} \mathbf{A}_{xx} \end{pmatrix} & \begin{pmatrix} \mathbf{A}_{xy} \end{pmatrix} & \begin{pmatrix} \mathbf{A}_{zx} \end{pmatrix} \\ \begin{pmatrix} \mathbf{A}_{xy} \end{pmatrix} & \begin{pmatrix} \mathbf{A}_{yy} \end{pmatrix} & \begin{pmatrix} \mathbf{A}_{yz} \end{pmatrix} \\ \begin{pmatrix} \mathbf{A}_{zx} \end{pmatrix} & \begin{pmatrix} \mathbf{A}_{yz} \end{pmatrix} & \begin{pmatrix} \mathbf{A}_{zz} \end{pmatrix} \end{pmatrix},$$

where $\mathbf{A}_{ij} = \hat{\mathbf{j}} \otimes \mathbf{p}_0 \times \hat{\mathbf{i}} + \hat{\mathbf{i}} \otimes \mathbf{p}_0 \times \hat{\mathbf{j}}, \; \forall \; \{i,j\} \in \{x,y,z\}$ and matching $\{\hat{\mathbf{i}}, \hat{\mathbf{j}}\} \in \{\hat{\mathbf{x}}, \hat{\mathbf{y}}, \hat{\mathbf{z}}\}$. Then the explicit form of the tensor yields

$$\mathbf{A}_s = \begin{pmatrix} \begin{pmatrix} 0 & 2p_{0z} & -2p_{0y} \\ 0 & 0 & 0 \\ 0 & 0 & 0 \end{pmatrix} & \begin{pmatrix} -p_{0z} & 0 & p_{0x} \\ 0 & p_{0z} & -p_{0y} \\ 0 & 0 & 0 \end{pmatrix} & \begin{pmatrix} p_{0y} & -p_{0x} & 0 \\ 0 & 0 & 0 \\ 0 & p_{0z} & -p_{0y} \end{pmatrix} \\ \begin{pmatrix} -p_{0z} & 0 & p_{0x} \\ 0 & p_{0z} & -p_{0y} \\ 0 & 0 & 0 \end{pmatrix} & \begin{pmatrix} 0 & 0 & 0 \\ -2p_{0z} & 0 & 2p_{0x} \\ 0 & 0 & 0 \end{pmatrix} & \begin{pmatrix} 0 & 0 & 0 \\ p_{0y} & -p_{0x} & 0 \\ -p_{0z} & 0 & p_{0x} \end{pmatrix} \\ \begin{pmatrix} p_{0y} & -p_{0x} & 0 \\ 0 & 0 & 0 \\ 0 & p_{0z} & -p_{0y} \end{pmatrix} & \begin{pmatrix} 0 & 0 & 0 \\ p_{0y} & -p_{0x} & 0 \\ -p_{0z} & 0 & p_{0x} \end{pmatrix} & \begin{pmatrix} 0 & 0 & 0 \\ 0 & 0 & 0 \\ 2p_{0y} & -2p_{0x} & 0 \end{pmatrix} \end{pmatrix}.$$

## Explicit Form of Tensor $\mathbf{T}_q$

The explicit terms of symmetric tensors $\mathbf{T}_q$ are shown in Table S1.

Table S1: Explicit vector terms of symmetric tensors $\mathbf{T}_q$

| ij | $(\mathbf{T}_x)_{ij}^{\mathrm{T}}$ | $(\mathbf{T}_y)_{ij}^{\mathrm{T}}$ | $(\mathbf{T}_z)_{ij}^{\mathrm{T}}$ |
|----|-----|-----|-----|
| xx | $(0, 2p_{0z}, -2p_{0y})$ | $(2p_{0z}, 0, 0)$ | $(-2p_{0y}, 0, 0)$ |
| yy | $(0, -2p_{0z}, 0)$ | $(-2p_{0z}, 0, 2p_{0x})$ | $(0, 2p_{0x}, 0)$ |
| zz | $(0, 0, 2p_{0y})$ | $(0, 0, -2p_{0x})$ | $(2p_{0y}, -2p_{0x}, 0)$ |
| xy | $(-2p_{0z}, 0, p_{0x})$ | $(0, 2p_{0z}, -p_{0y})$ | $(p_{0x}, -p_{0y}, 0)$ |
| yz | $(0, p_{0y}, -p_{0z})$ | $(p_{0y}, -2p_{0x}, 0)$ | $(-p_{0z}, 0, 2p_{0x})$ |
| zx | $(2p_{0y}, -p_{0x}, 0)$ | $(-p_{0x}, 0, p_{0z})$ | $(0, p_{0z}, -2p_{0y})$ |

## Explicit Form of the Trace Derivatives $\partial_q t_2$ and $\partial_q t_4$

The explicit identities for the partial derivatives $\partial_q t_2$ with $q \in \{x, y, z\}$ are defined as

$$\begin{aligned}
\partial_q t_2 &= 2\Re\left(\operatorname{Tr}\left[\mathbf{Q}_{\mathrm{pm}}^* \cdot \partial_q \mathbf{\Delta}_{\mathrm{pm}}\right]\right) \\
&= \frac{2}{3}\Re\left(\operatorname{Tr}\left[\mathbf{Q}_{\mathrm{pm}}^* \cdot (\mathbf{N} \cdot \hat{\mathbf{q}}) + \mathbf{Q}_{\mathrm{pm}}^* \cdot \left(\frac{s}{2}\mathbf{T}_q \cdot \mathbf{d}\right)\right]\right),
\end{aligned} \quad (S40)$$

$$\begin{aligned}
\partial_q t_4 &= 2\Re\left(\operatorname{Tr}\left[\mathbf{\Delta}_{\mathrm{pm}}^* \cdot \partial_q \mathbf{\Delta}_{\mathrm{pm}}\right]\right) \\
&= \frac{2}{9}\Re\Bigg[\operatorname{Tr}\bigg((\mathbf{N}^* \cdot \mathbf{d}) \cdot (\mathbf{N} \cdot \hat{\mathbf{q}}) + \frac{s}{2}(\mathbf{N}^* \cdot \mathbf{d}) \cdot (\mathbf{T}_q \cdot \mathbf{d}) \\
&\quad + \frac{s^*}{2}[(\mathbf{A}_{\mathrm{s}}^* \cdot \mathbf{d}) \cdot \mathbf{d}] \cdot (\mathbf{N} \cdot \hat{\mathbf{q}}) + \frac{s}{2}[(\mathbf{A}_{\mathrm{s}}^* \cdot \mathbf{d}) \cdot \mathbf{d}] \cdot (\mathbf{T}_q \cdot \mathbf{d})\bigg)\Bigg] \\
&= \frac{1}{9}\Re\left(\operatorname{Tr}\left[2(\mathbf{N}^* \cdot \mathbf{d}) \cdot (\mathbf{N} \cdot \hat{\mathbf{q}}) + s(\mathbf{N}^* \cdot \mathbf{d}) \cdot (\mathbf{T}_q \cdot \mathbf{d})\right]\right) \\
&\quad + \frac{1}{18}\Re\left(\operatorname{Tr}\left[2s^*((\mathbf{A}_{\mathrm{s}}^* \cdot \mathbf{d}) \cdot \mathbf{d}) \cdot (\mathbf{N} \cdot \hat{\mathbf{q}}) + |s|^2((\mathbf{A}_{\mathrm{s}}^* \cdot \mathbf{d}) \cdot \mathbf{d}) \cdot (\mathbf{T}_q \cdot \mathbf{d})\right]\right).
\end{aligned} \quad (S41)$$

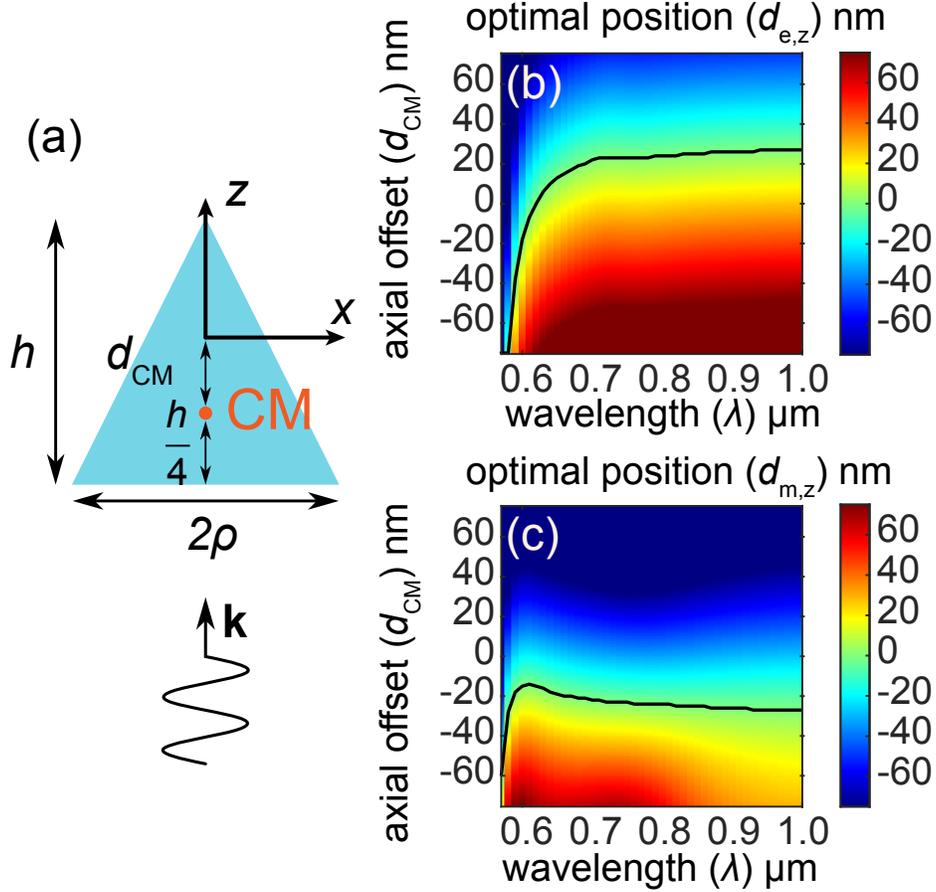

Figure S2: Verification of the dielectric cone OSCs. (a) The panel duplicates Fig. 2a, and is included for reference. A dielectric cone with $\varepsilon_{\mathrm{r}} = 10$ and $h = 2\rho = 300\,\mathrm{nm}$ is illuminated by an $x$-polarized plane wave. A local Cartesian coordinate system is shifted by $d_{\mathrm{CM}}$ from the CM along the $z$-axis. (b) and (c) respectively show the pseudocolor maps of the optimal positions of the electric and magnetic expansion centers in nm. The black solid lines indicate the OSC positions obtained from $d_{\mathrm{e},z}$ and $d_{\mathrm{m},z}$ that perfectly trace the color-map levels of $d_{\mathrm{e},z} = 0$ and $d_{\mathrm{m},z} = 0$.

## Independence of the OSCs on the Initial Reference Point

Figure S2 illustrates the independence of the OSC positions on the position of the initial reference point. The figure employs the geometry and parameters of Fig. 2a of the main text. The specific aim here is to confirm that *obtaining the spatial multipolar spectra for every value of $d_{CM}$ is not necessary and could be chosen assuming convenience and symmetry.*

Panels (b) and (c) in Fig. S2 present a numerical confirmation that it is sufficient to compute $d_{\mathrm{e/m},z}$ for any convenient choice of $d_{\mathrm{CM}}$ (e.g, $d_{\mathrm{CM}} = 0$), and that any fixed reference point around the CM can directly provide the required positions of the electric and magnetic OSCs. We may want to recapitulate that choosing the CM for the initial reference point and aligning the coordinate system with the axis of symmetry (for axisymmetric cases) could lead to improved accuracy. Once the initial reference system is chosen, the electric and magnetic multipoles are separately computed at the positions of the electric and magnetic OSCs using the analytical formulas of Tables 2 and 4 along with, for instance, $\mathbf{d} = d_{e,z}\hat{\mathbf{z}}$. So there is no need to perform additional, time-costly numerical multipole re-expansions from the field data.